\documentclass{elsart}
\usepackage{psfrag,graphicx,float,array}
\usepackage{amsmath,amssymb}
\journal{Nuclear Physics B}
\usepackage[mathscr]{eucal}
\newcommand{\al}{\alpha}
\newcommand{\be}{\beta}
\newcommand{\de}{\delta}
\newcommand{\ep}{\epsilon}
\newcommand{\vep}{\varepsilon}
\newcommand{\ga}{\gamma}

\newcommand{\la}{\lambda}
\newcommand{\om}{\omega}
\newcommand{\si}{\sigma}

\newcommand{\vp}{\varphi}
\newcommand{\vt}{\vartheta}
\newcommand{\ze}{\zeta}
\newcommand\Om\Omega
\newcommand{\De}{\Delta}
\newcommand{\Ga}{\Gamma}
\newcommand{\La}{\Lambda}
\newcommand{\Si}{\Sigma}
\newcommand{\bs}{\mathbf{s}}
\newcommand{\bx}{\mathbf{x}}
\newcommand{\bnu}{\boldsymbol{\nu}}

\newcommand{\bxi}{\boldsymbol{\xi}}
\newcommand{\bS}{\mathbf{S}}
\newcommand{\bu}{\mathbf{u}}
\newcommand{\bv}{\mathbf{v}}
\newcommand{\bw}{\mathbf{w}}

\newcommand{\tPhi}{\widetilde{\Phi}}
\newcommand\tPsi{\widetilde\Psi}
\newcommand{\hPhi}{\widehat{\Phi}}
\newcommand{\hPsi}{\widehat{\Psi}}
\def\RR{\mathbb{R}}

\newcommand{\cA}{{\mathcal A}}

\newcommand{\cE}{{\mathcal E}}
\newcommand{\cF}{{\mathcal F}}

\newcommand{\cH}{{\mathcal H}}

\newcommand{\cN}{{\mathcal N}}

\newcommand{\cZ}{{\mathcal Z}}
\def\Bx{\bar x}
\def\Bxi{\bar\xi}

\def\BcE{\,\overline{\!\cE}{}}

\newcommand{\lf}{\lfloor}
\newcommand{\rf}{\rfloor}
\newcommand{\pa}{\partial}
\newcommand{\ra}{\to}

\def\ket#1{|#1\rangle}

\let\ds\displaystyle

\newcommand{\ms}{\mspace{1mu}}
\newcommand{\mms}{\mspace{2mu}}
\renewcommand{\leq}{\leqslant}
\renewcommand{\geq}{\geqslant}
\renewcommand{\leq}{\leqslant}
\renewcommand{\geq}{\geqslant}
\newcommand{\erf}{\operatorname{erf}}

\newcommand{\tr}{\operatorname{tr}}

\newcommand{\iu}{{\mathrm i}}
\newcommand{\pd}{\partial}

\DeclareMathOperator\Imag{Im}
\newcommand\Hsc{H_{\mathrm{sc}}}
\newcommand\Esc{E_{\mathrm{sc}}}
\newcommand\Zsc{Z_{\mathrm{sc}}}

\begin{document}
\begin{frontmatter}
\title{A novel quasi-exactly solvable spin chain with nearest-neighbors interactions}
\author{A. Enciso}, \author{F. Finkel}, \author{A.
  Gonz{\'a}lez-L{\'o}pez}, \author{M.A. Rodr{\'\i}guez}
\address{Depto.~de F{\'\i}sica Te{\'o}rica II, Universidad
Complutense, 28040 Madrid, Spain}
\date{April 23, 2007}

\begin{abstract}
  In this paper we study a novel spin chain with nearest-neighbors
  interactions depending on the sites coordinates, which in some sense is
  intermediate between the Heisenberg chain and the spin chains of
  Haldane--Shastry type. We show that when the number of spins is sufficiently
  large both the density of sites and the strength of the interaction between
  consecutive spins follow the Gaussian law. We develop an extension of the
  standard freezing trick argument that enables us to exactly compute a
  certain number of eigenvalues and their corresponding eigenfunctions. The
  eigenvalues thus computed are all integers, and in fact our numerical
  studies evidence that these are the only integer eigenvalues of the chain
  under consideration. This fact suggests that this chain can be regarded as a
  finite-dimensional analog of the class of quasi-exactly solvable
  Schr\"odinger operators, which has been extensively studied in the last two
  decades. We have applied the method of moments to study some statistical
  properties of the chain's spectrum, showing in particular that the density
  of eigenvalues follows a Wigner-like law. Finally, we emphasize that, unlike
  the original freezing trick, the extension thereof developed in this paper
  can be applied to spin chains whose associated dynamical spin model is only
  quasi-exactly solvable.
\end{abstract}

\end{frontmatter}

\section{Introduction}\label{S:Intro}

Solvable spin chains have enjoyed a growing popularity in the last few years, due in part to
their novel applications to SUSY Yang--Mills and string
theories~\cite{MZ03,BC04,RV04,BS05,FKM05,Go05}. The prime example of
such chains is the celebrated Heisenberg model describing $N$
spins in a one-dimensional lattice with isotropic nearest-neighbors interactions
independent of the site. The Hamiltonian of the model is given by~\cite{He28}
\begin{equation}\label{He}
\cH_{\mathrm{He}}=\sum_i \bS_i\cdot\bS_{i+1}\,,
\end{equation}
where $\bS_i=(S_i^x,S_i^y,S_i^z)$ is the spin operator of the $i$-th site,
the sum runs from $1$ to $N$ (as always hereafter), and $\bS_{N+1}=\bS_1$. As is
well-known, for spin $1/2$ the model~\eqref{He} can be exactly solved using
Bethe's ansatz~\cite{Be31,Hu38,CP62}. Several (partially) solvable
generalizations of the Heisenberg chain~\eqref{He} with short range interactions (at most
between next to nearest neighbors) have been subsequently proposed
in the literature. These include, in particular, the family of chains with
arbitrary spin and nearest-neighbors interactions polynomial in
$\bS_i\cdot\bS_{i+1}$ of Refs.~\cite{Ba82,Ta82}, as well as several models whose
ground state can be written in terms of ``valence bonds''~\cite{MG69,AKLT87}.

A different type of solvable spin chain with long-range position-dependent
couplings was introduced independently by Haldane~\cite{Ha88} and
Shastry~\cite{Sh88}. This chain describes a system of $N$ spins equally spaced
on a circle, such that the strength of the interaction between each pair of
spins is inversely proportional to their chord distance.
The motivation for introducing the HS chain~\eqref{HS} was the fact that its exact
ground state coincides with Gutzwiller's variational
wave function for the Hubbard model~\cite{Hu63,Gu63,GV87}
when the strength of the on-site interaction tends to infinity.
We shall write the
Haldane--Shastry (HS) Hamiltonian as
\begin{equation}\label{HS}
\cH_{\mathrm{HS}}=\frac12\sum_{i<j} \sin(\vt_i-\vt_j)^{-2}(1-S_{ij})\,,\qquad\vt_i\equiv\frac{i\pi}N\,,
\end{equation}
where $S_{ij}$ is the operator exchanging the $i$-th and $j$-th
spins. Although the particles' spin in the original HS chain was
assumed to be $1/2$, one can more generally consider particles
with $n$ internal degrees of freedom transforming under the
fundamental representation of $\mathrm{su}(n)$. In this case the
spin permutation operators can be written in terms of the
fundamental $\mathrm{su}(n)$ generators $J^\al_k$ at each site $k$
(normalized such that $\tr(J^\al_kJ^\be_k)=\frac12\,\de^{\al\be}$)
as
\[
S_{ij}=\frac1n+2\sum_{\al=1}^{n^2-1}J_i^\al J_j^\al\,.
\]
Note that for spin $1/2$ particles ($n=2$), we have $\bS_i=(J_i^1,J_i^2,J_i^3)$.

The HS chain is naturally related to the scalar Sutherland model of $A_N$ type
\cite{Su71,Su72} and its spin version introduced in
Refs.~\cite{HH92,HW93,MP93}. In fact, Polychronakos~\cite{Po93} noted that the
complete integrability of the HS chain could be deduced from that of the spin
Sutherland model by suitably taking the strong coupling limit (the so-called
``freezing trick''). Moreover, the latter author applied this technique to
construct an integrable spin chain related to the Calogero (rational) model
of $A_N$ type~\cite{Ca71}. The Hamiltonian of this chain, usually referred to
in the literature as the Polychronakos--Frahm (PF) chain, reads
\begin{align}\label{PF}
\cH_{\mathrm{PF}}&=\sum_{i<j}(\ze_i-\ze_j)^{-2}\,(1-S_{ij})\,,
\end{align}
where the chain sites $\ze_i$ are now the equilibrium positions of $N$
particles in the scalar part of the potential of the Calogero spin model of
$A_N$ type. In particular, the sites of the PF chain are not equally spaced,
unlike those of the HS chain. Indeed, Frahm~\cite{Fr93} pointed out that the
sites of the PF chain are the zeros of the $N$-th Hermite polynomial, which
satisfy the system of algebraic equations
\begin{equation}\label{Hermite0s}
\ze_i=\sum_{j\neq i}\frac1{\ze_i-\ze_j}\,,\qquad i=1,\dots,N\,.
\end{equation}

It turns out that both the HS and the PF chains, featuring long-range
position-dependent interactions, can be solved in a more detailed and explicit way than the
chains of Heisenberg type, characterized by the short range and position independence
of the interactions. For instance, the partition function of the
models~\eqref{HS} and~\eqref{PF} can be evaluated in closed form for arbitrary $N$
using the freezing trick~\cite{Po94,FG05}, and its expression is relatively simple
in both cases. The spectrum, which consists of a set of integers (consecutive in
the case of the PF chain), is highly degenerate due to an underlying
Yangian symmetry~\cite{BGHP93}. Moreover, it has been recently shown that even for moderately large
$N$ the density of eigenvalues of the HS chain is Gaussian to a high degree of approximation, and that
the density of spacings between consecutive levels follows a simple distribution
different from the usual Poisson or Wigner laws~\cite{FG05}. In fact, there is
strong evidence that these results also hold for other spin chains of Haldane--Shastry type,
as e.g.~the $BC_N$ or the supersymmetric versions of the original HS
chain~\cite{EFGR05,BB06}.

In this paper we shall consider a novel type of spin chain, which in some sense is
intermediate between the Heisenberg and the Polychronakos--Frahm chains. Its Hamiltonian
is obtained from that of the PF chain~\eqref{PF}-\eqref{Hermite0s} by retaining only
nearest-neighbors interactions, namely
\begin{align}\label{cH}
\cH&=\sum_i(\xi_i-\xi_{i+1})^{-2}\,(1-S_{i,i+1})\,,
\end{align}
where the sites $\xi_i$ are defined by restricting the sum in
Eq.~\eqref{Hermite0s} to nearest neighbors:
\begin{equation}\label{sites}
\xi_i=\frac1{\xi_i-\xi_{i-1}}+\frac1{\xi_i-\xi_{i+1}}\,,\qquad i=1,\dots,N\,.
\end{equation}
In the previous equations we are identifying $S_{N,N+1}$ with $S_{N1}$,
$\xi_{N+1}$ with~$\xi_1$ and $\xi_0$ with $\xi_N$.

We shall see that the chain~\eqref{cH} possesses several
remarkable properties, whose study is the purpose of this paper.
In the first place, the spin chain~\eqref{cH} is related along the lines of the
freezing trick to the spin dynamical model
\begin{multline}\label{H}
H=-\sum_i\pa_{x_i}^2+a^2
r^2+\sum_i\frac{2a^2}{(x_i-x_{i-1})(x_i-x_{i+1})}\\
+\sum_i\frac{2a}{(x_i-x_{i+1})^2}\,(a-S_{i,i+1})\,,
\end{multline}
where $r^2=\sum_i x_i^2$, $a>1/2$, and we have identified
$x_0\equiv x_N$ and $x_{N+1}\equiv x_1$. We have shown in our recent
papers~\cite{EFGR05b,EFGR07} that an infinite proper subset of the
spectrum of the Hamiltonian~\eqref{H}
can be computed in closed form, so that this model is
\emph{quasi-exactly solvable} (QES)~\cite{Tu88,Sh89,Us94}.
The Hamiltonian~\eqref{H} is a spin version of the QES scalar model
\begin{equation}\label{Hsc}
\Hsc=H|_{S_{i,i+1}\ra1}
\end{equation}
introduced in~\cite{AJK01} by Auberson, Jain and Khare.
In particular, it was shown in the latter reference that
\begin{equation}\label{mu}
\mu=\e^{-\frac a2\ms r^2}\prod_i |x_i-x_{i+1}|^a
\end{equation}
is the ground state function of the model~\eqref{Hsc},
with eigenvalue $E_0=Na(2a+1)$.
We shall prove that the sites $\xi_i$ of the chain~\eqref{cH} are in fact
the coordinates of the unique maximum of $\mu$ in the domain
\begin{equation}\label{C}
C=\big\{\bx\in\RR^N\mid x_1<\cdots<x_N\big\}.
\end{equation}
By numerically solving Eq.~\eqref{sites}, we
shall see that for sufficiently large $N$ the chain sites
are normally distributed with zero mean and unit variance.
We shall present a simple deduction of this property based
on the analysis of the continuous limit of the algebraic system~\eqref{sites}
as $N\to\infty$. As a byproduct, we shall obtain an analytic formula
providing a very accurate approximation to the sites' coordinates,
valid even for moderately large values of $N$. Another nontrivial
consequence of this formula is the fact that
for large $N$ the strength of the coupling between the
spins $i$ and $i+1$ as a function of their mean coordinate
$(\xi_i+\xi_{i+1})/2$ also follows the Gaussian law, but with zero mean
and variance $1/2$.

The spectral properties of the spin chain Hamiltonian~\eqref{cH} are also
remarkable. Indeed, by a suitable modification of the freezing trick one can
show that $\cH$ possesses the eigenvalues $0,1,2$ for arbitrary values of $N$
and~$n$, and exactly compute their corresponding eigenstates. Our numerical
simulations evidence that these energies are the three lowest ones, and that
for spin $1/2$ none of the remaining eigenvalues of $\cH$ are
integers\footnote{%
  As a matter of fact, the previous assertion does not hold for the cases
  $N=3$ (for which the chain~\eqref{cH} reduces to the PF chain, whose
  eigenvalues are known to be integers), and $N=4$ (for which our numerical
  simulations indicate that all the eigenvalues are also integers). Therefore,
  in the rest of the paper we shall exclude these special cases from our
  discussion.}. For $n>2$, the spectrum of $\cH$ also contains the integer
eigenvalue $3$, which appears to be not the fourth but the fifth lowest
energy. The above properties suggest that the model~\eqref{cH} could be
regarded as a quasi-exactly solvable chain, in the sense that only a certain
number of eigenvalues and their corresponding eigenvectors of the Hamiltonian
$\cH$ can be computed in closed form.

We have also studied the distribution of energy levels of the chain~\eqref{cH}
for a large number of particles. Since the partition function of this chain is
not known, we have performed a numerical calculation of the density of levels
using the methods of moments~\cite{BRP92,AFGLM01}. It turns out that, in
contrast with the typical behavior of spin chains of Haldane--Shastry type,
the distribution of levels clearly deviates from the Gaussian law. As an
indication of the accuracy of the approximate level density derived via the
moments method, we have compared its mean and variance with the exact values
obtained by taking traces of suitable powers of the Hamiltonian. {}From this
discussion it also follows that for large $N$ the mean and variance of the
energy behave as $N^3$ and $N^5$, respectively, just as for the trigonometric
chains of HS type in Refs.~\cite{FG05,EFGR05}.

The paper is organized as follows. In Section~\ref{S:Sites} we study the
distributions of the chain sites and the couplings, comparing the results
obtained with those for the PF chain. Section~\ref{S:QES}, which is the core
of the paper, is devoted to the determination of the integer eigenvalues of
the chain~\eqref{cH} and their corresponding eigenstates. We also present in
this section a detailed example for the case of $5$ particles of spin $1$,
which motivates a number of conjectures on the degeneracy of the integer
levels. In Section~\ref{S:stas} we use the method of moments to approximately
compute the density of levels of the chain~$\cH$, showing that it follows a
Wigner-like law. Finally, in Section~\ref{S:conc} we summarize our conclusions
and outline possible future developments. For the reader's convenience, in
Appendix~\ref{appA} we present some background material on the exact
eigenfunctions of the spin dynamical model~\eqref{H} used in
Section~\ref{S:QES}, while in Appendix~\ref{appB} we include an overview of
the method of moments.

\section{The chain sites}\label{S:Sites}

We shall start this section by proving that the sites of the
chain~\eqref{cH} are the coordinates of the unique maximum in the domain~\eqref{C} of the
ground state function~\eqref{mu} of the scalar Hamiltonian~\eqref{Hsc}.
It is convenient to write the ground state as
\[
\mu=\e^{a\ms\la(\bx)}\,,
\]
where
\[
\la(\bx)=\sum_i\log|x_i-x_{i+1}|-\frac{r^2}2
\]
has the same extrema as $\mu$ and is independent of $a$.
Thus the equations~\eqref{sites} defining the chain sites are just
the conditions for $\bxi=(\xi_1,\dots,\xi_N)$ to be a critical
point of $\la$. The existence of a maximum of $\la$ in $C$ is clear,
since it is continuous in $C$ and tends to $-\infty$ both on its boundary
and as $r\to\infty$. Uniqueness follows from the fact that the Hessian of $\la$ is negative
definite in $C$. Indeed, by Gerschgorin's theorem~\cite[15.814]{GR00},
the eigenvalues of the Hessian of $\la$ at $\bx$ lie in the union of the intervals
\[
\Big[\,\frac{\pd^2 \la}{\pd x_i^2}-\gamma_i\,,\frac{\pd^2 \la}{\pd
x_i^2}+\gamma_i\,\Big]\,, \quad\text{where } \gamma_i=\sum_{j\neq
i}\Big|\,\frac{\pd^2 \la}{\pd x_i\pd x_j}\,\Big|\,.
\]
Since
\begin{gather*}
\frac{\pd^2\la}{\pd x_i^2}=-1-(x_i-x_{i+1})^{-2}-(x_i-x_{i-1})^{-2},\\[1mm]
\frac{\pd^2\la}{\pd x_i\pd x_{i\pm1}}=(x_i-x_{i\pm1})^{-2},\qquad
\frac{\pd^2\la}{\pd x_i\pd x_{j}}=0\;\text{ if }\;j\neq i,i\pm1,
\end{gather*}
we have
\[
\frac{\pd^2 \la}{\pa x_i^2}+\ga_i=-1\,,
\]
and thus all the eigenvalues of the Hessian of $\la$ are strictly negative.

Since $\la$ and $C$ are invariant under the transformation
\[
x_i\mapsto -x_{N-i+1}\,,\qquad i=1,\dots, N\,,
\]
and $\la$ has a unique maximum in $C$, it follows that
\begin{equation}\label{symxi}
\xi_i=-\xi_{N-i+1},
\end{equation}
so that the chain sites are symmetric about the origin.
In particular, the center of mass of the spins vanishes, i.e.,
\begin{equation}\label{xi=0}
\Bxi\equiv\frac1N\sum_i\xi_i=0\,.
\end{equation}

\begin{figure}[h]
\begin{center}
\psfrag{x}{\footnotesize $\xi_i$}
\includegraphics[height=6cm]{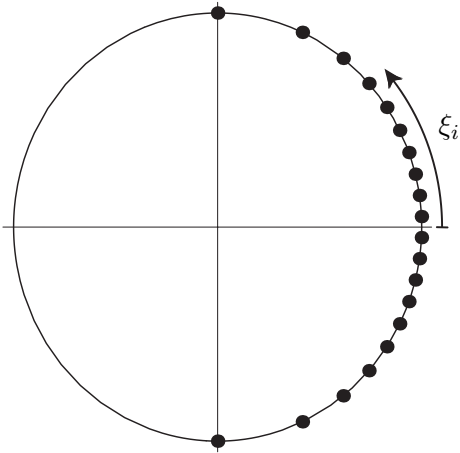}
\begin{quote}
\caption{Sites of the chain $\cH$ for $N=20$ spins.\label{fig:sites20}}
\end{quote}
\end{center}
\end{figure}

We have numerically solved equations~\eqref{sites} for the chain sites for up
to 250 spins. Before presenting our conclusions, a remark on the configuration
of these sites is in order. The attentive reader might have been surprised by
our claim that the chain~\eqref{cH} features only nearest-neighbors
interactions, in spite of the fact that the first spin interacts with the last
one. We can avoid this objection by regarding the site coordinate $\xi_i$ as an
arc length in a circle of radius $2\xi_N/\pi$, see Fig.~\ref{fig:sites20}. In
this way the spins at the sites $\xi_1$ and $\xi_N$ are indeed nearest-neighbors
and, moreover, the strength of the interactions are inversely proportional to
the squared distance between consecutive spins, measured along the arc.

\begin{figure}[h]
\begin{center}
\psfrag{F}[Bc][Bc][1][0]{\footnotesize\,$\cF(x),F(x)$}
\psfrag{x}{\footnotesize $x$}
\includegraphics[height=6cm]{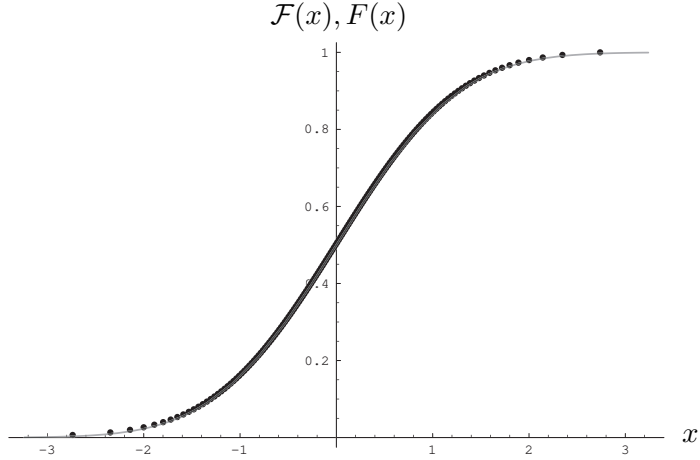}
\begin{quote}
\caption{Cumulative distribution functions $\cF(x)$ (at its discontinuity points)
and $F(x)$ (continuous grey line) for $N=150$ spins.\label{fig:sitesdist150}}
\end{quote}
\end{center}
\end{figure}

It is apparent from Fig.~\ref{fig:sites20}
(and also follows immediately from Eq.~\eqref{sites}) that the sites $\xi_i$ are not equally spaced.
In fact, our computations show that for large values of $N$ the sites $\xi_i$
follow with great accuracy a Gaussian distribution with zero mean and unit variance.
More exactly, the cumulative density of sites (normalized to unity)
\begin{equation}\label{cF}
\cF(x)=N^{-1}\sum_i\theta(x-\xi_i)\,,
\end{equation}
where $\theta$ is Heaviside's step function, is approximately given by
\begin{equation}\label{F}
F(x)=\frac12\,\Big[1+\erf\!\big(x/\sqrt2\ms\big)\Big]\,.
\end{equation}
The agreement between the functions $\cF$ and $F$ is remarkably
good for $N\gtrsim100$ (see~Fig.~\ref{fig:sitesdist150} for the
case $N=150$) and increases steadily with $N$, e.g., the mean
square error of the fit for $100$, $150$ and 200 spins are
respectively $2.6\times10^{-5}$, $1.1\times10^{-5}$ and $7.9\times 10^{-6}$.

The fact that for large $N$ the cumulative density of sites is well approximated by the Gaussian
law~\eqref{F} can be justified by the following heuristic argument. Let $x(t,N)$
be a smooth function such that $x(i,N)=\xi_i$ for $i=1,\dots,N$, and define
the rescaled function $y(s,\ep)=x(s/\ep,1/\ep)$. By Eq.~\eqref{sites}, the latter function
must satisfy the relation
\begin{equation}\label{sitesy}
\frac1{y(s,\ep)-y(s-\ep,\ep)}+\frac1{y(s,\ep)-y(s+\ep,\ep)}=y(s,\ep)
\end{equation}
for $\ep=1/N\ll 1$ and $s=\frac1N,\frac2N,\dots,1$. Let us now assume that
Eq.~\eqref{sitesy} holds for all $s\in\RR$ and all $\ep\ll 1$. Writing
\[
y(s,\ep)=\sum_{k=0}^\infty y_k(s)\ep^k\,,
\]
and using the expansion
\[
y(s,\ep)-y(s\pm\ep,\ep)=\mp\,y'_0(s)\ep-\Big(\frac{y''_0(s)}2\pm y'_1(s)\Big)\ep^2+O(\ep^3)
\]
the leading term in Eq.~\eqref{sitesy} yields the differential equation
\[
y_0''=y_0\,{y_0'}^2\,.
\]
The general solution of this equation is implicitly given by
\begin{equation}\label{sy0}
s=c_0+c_1\erf\!\big(y_0(s)/\sqrt2\ms\big)\,.
\end{equation}
Hence, up to terms of order $\ep=1/N$, the cumulative distribution function
of the chain sites (normalized to unity) is approximated by the continuous function
\[
F(x)=c_0+c_1\erf\!\big(x/\sqrt2\ms\big)\,.
\]
The normalization conditions $F(-\infty)=0$ and $F(\infty)=1$ imply that $c_0=c_1=1/2$,
and thus the empiric law~\eqref{F} is recovered.

\begin{figure}[h]
\begin{center}
\psfrag{x}{\footnotesize$\xi_k$}
\psfrag{k}{\footnotesize $k$}
\includegraphics[height=6cm]{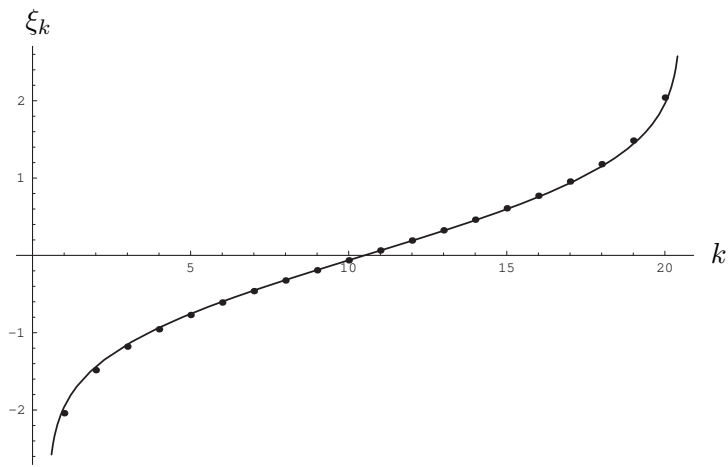}
\begin{quote}
\caption{Sites coordinates $\xi_k$ and their continuous approximation~\eqref{xigood}
for $N=20$ spins.\label{fig:xis20}}
\end{quote}
\end{center}
\end{figure}

{}From Eq.~\eqref{sy0} (with $c_0=c_1=1/2$) it follows that the site $\xi_k$ can be
determined up to terms of order $1/N$ by the formula
\begin{equation}\label{xibad}
\xi_k\simeq\sqrt2\,\erf^{-1}\Big(\frac{2k-N}N\Big)\,.
\end{equation}
If the sites $\xi_k$ were exactly given by the previous formula, they would satisfy the identity
\[
\erf\!\big(\xi_k/\sqrt 2\big)+\erf\!\big(\xi_{N-k+1}/\sqrt 2\big)=\frac2N\,,
\]
which is clearly inconsistent with the exact relation~\eqref{symxi}.
However, the slightly modified formula
\begin{equation}\label{xigood}
\xi_k\simeq\sqrt2\,\erf^{-1}\Big(\frac{2k-N-1}N\Big)
\end{equation}
differs from~\eqref{xibad} by a term of order $1/N$ and is fully consistent
with the relation~\eqref{symxi}.
Although both~\eqref{xibad} and~\eqref{xigood} provide an excellent approximation to the chain sites
for large $N$, the latter equation is always more accurate than the former, and can be used to estimate $\xi_k$
with remarkable precision even for relatively low values of $N$, cf.~Fig.~\ref{fig:xis20}.

\begin{figure}[h]
\begin{center}
\psfrag{x}[Bc][Bc][1][0]{\footnotesize $\xi_N$}
\psfrag{N}{\footnotesize $N$}
\includegraphics[height=6cm]{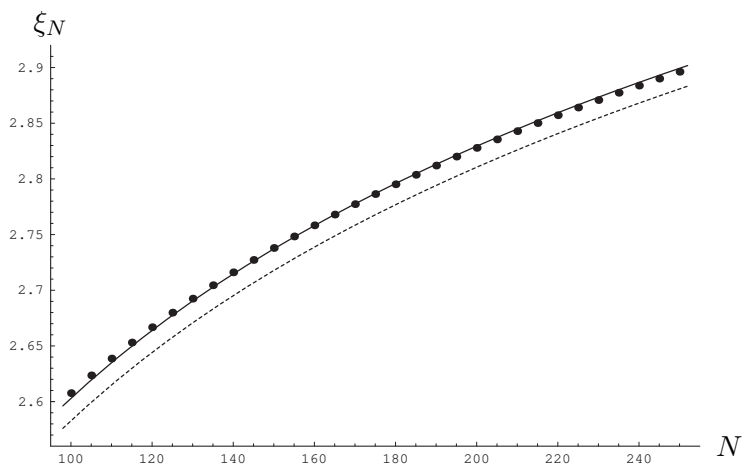}
\begin{quote}
\caption{Position of the last spin $\xi_N$ for $N=100,105,\dots,250$
and its continuous approximation~\eqref{xiNsimpler}-\eqref{etaalfa}
for $\al=0.94$ (solid line) and $\al=1$ (dashed line).\label{fig:xmax100}}
\end{quote}
\end{center}
\end{figure}

It is also of interest to determine whether the position of the last spin
tends to infinity as $N\to\infty$, since according to our interpretation of the
chain's geometry the number $2\xi_N/\pi$ is the radius of the circle on which
the spins lie. {}From Eq.~\eqref{xigood} it follows that for large $N$ the
last spin's coordinate $\xi_N$ is approximately given by
\begin{equation}\label{xiN}
\xi_N\simeq\sqrt2\,\erf^{-1}\Big(1-\frac1N\Big)\,,
\end{equation}
so that $\xi_N$ should diverge as $N\to\infty$. Of course, this
assertion should be taken with some caution, since in Eq.~\eqref{xigood}
the argument of the inverse error function is correct only up to terms of order $1/N$.
In order to check the correctness of the approximation~\eqref{xiN},
we recall the asymptotic expansion of
$\erf^{-1}(u)$ for $u\to 1$ in Ref.~\cite{BEJ76} to replace~\eqref{xiN} by the simpler formula
\begin{equation}\label{xiNsimpler}
\xi_N\simeq\sqrt{2\eta-\log\eta}\,,
\end{equation}
where
\[
\eta=\log\Big(\frac N{\sqrt\pi}\Big)\,.
\]
As can be seen in Fig.~\ref{fig:xmax100}, the approximate formula~\eqref{xiNsimpler}
qualitatively reproduces the growth of $\xi_N$ when $N$ ranges from $100$ to $250$.
A greater accuracy can be achieved by introducing
an adjustable parameter in Eq.~\eqref{xiN} through the
replacement $1/N\to\al/N$,
so that in Eq.~\eqref{xiNsimpler} $\eta$ becomes
\begin{equation}\label{etaalfa}
\eta=\log\Big(\frac N{\al\sqrt\pi}\Big)\,.
\end{equation}
In Fig.~\ref{fig:xmax100}, we have also plotted the law~\eqref{xiNsimpler}-\eqref{etaalfa}
with the optimal value $\al=0.94$, which is in excellent
agreement with the numerical values of $\xi_N$ for $N=100,105,\dots,250$.

\begin{figure}[h]
\begin{center}
\psfrag{X}[Bc][Bc][1][0]{\footnotesize $h_k$}
\psfrag{x}{\footnotesize $\Bxi_k$}
\includegraphics[height=6cm]{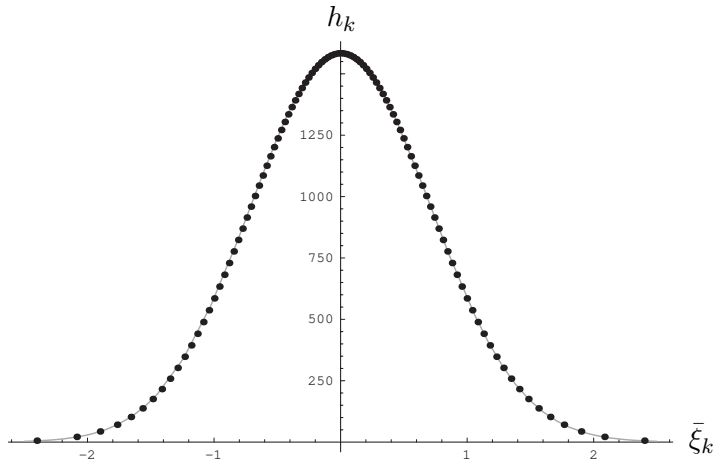}
\begin{quote}
\caption{Plot of the coupling between the spins $k$ and $k+1$ versus their mean position,
fitted by the Gaussian~\eqref{couplings},
for $N=100$ spins.\label{fig:couplings100}}
\end{quote}
\end{center}
\end{figure}

The last property of the spin chain~\eqref{cH} that we shall analyze in this section
is the dependence of the coupling between neighboring spins on their mean coordinate. Calling
\begin{equation}
  h_k=(\xi_k-\xi_{k+1})^{-2}\,,\qquad\Bxi_k=\frac{\xi_k+\xi_{k+1}}2\,,
  \label{hk}
\end{equation}
we shall now see that when $N\gtrsim100$ the Gaussian law
\begin{equation}\label{couplings}
h_k\simeq\frac{N^2}{2\pi}\,\e^{-{\Bxi_k}^{\ms 2}}
\end{equation}
holds with remarkable precision, cf.~Fig.~\ref{fig:couplings100}. Indeed, if $x=x(k)$ denotes the RHS
of Eq.~\eqref{xigood} we have
\[
2k=N\erf\Big(\frac x{\sqrt2}\Big)+N+1\,,
\]
so that
\begin{equation}\label{dxdk}
\frac{\d x}{\d k}=\frac{\sqrt{2\pi}}N\,\e^{\frac12x^2}
\end{equation}
is of order $1/N$. Hence, up to terms of order $1/N$ we have
\begin{equation}\label{precouplings}
\Big[x\Big(k-\frac12\Big)-x\Big(k+\frac12\Big)\Big]^{-2}\simeq
\Big(\frac{\d x}{\d k}\Big)^{-2}=\frac{N^2}{2\pi}\,\e^{-x^2(k)}\,.
\end{equation}
Since (up to terms of order $1/N^2$)
\[
x(k)\simeq\frac12\,\Big[x\Big(k-\frac12\Big)+x\Big(k+\frac12\Big)\Big]\,,
\]
Eq.~\eqref{couplings} follows from~\eqref{precouplings} replacing $k$ by $k+\frac12$.

We shall finish this section with a brief comparison of the previous
properties with those of the PF chain~\eqref{PF}. For large $N$, the density
of sites of the PF chain (normalized to unity), that is the density of zeros
of the $N$-th Hermite polynomial, is asymptotically given by the circular law~\cite{CP78b}
\begin{equation}\label{rhoPF}
\rho_N(x)=\frac1{\pi N}\,\sqrt{2N-x^2}\,.
\end{equation}
The last site $\ze_N$ of the PF chain grows with $N$ much faster
than the corresponding site $\xi_N$ of the chain~\eqref{cH}, for the largest zero
of the $N$-th Hermite polynomial behaves as
$\sqrt{2N}+O(N^{-1/6})$; see, e.g., \cite{Do06}.
Finally, in contrast with the PF chain, the polynomials
determined for each $N\geq 2$ by the corresponding sites $\xi_i$ of the chain~\eqref{cH}
do not form an orthogonal family. In other words, the polynomials
\begin{equation}\label{pN}
p^{(N)}(z)\equiv\prod_i(z-\xi_i)
\end{equation}
do not satisfy a three-term recursion relation of the form
\[
p^{(N+1)}(z)=z p^{(N)}(z)-a_N p^{(N-1)}(z)\,,
\]
as it can be verified using the explicit expressions
\[
p^{(2)}(z)=z^2-1\,,\qquad p^{(3)}(z)=z^3-\frac{3z}2\,,\qquad p^{(4)}(z)=z^4-2 z^2+\frac14\,.
\]

\section{Quasi-exact solvability of the spin chain}\label{S:QES}

In this section we shall see that the spin chain~\eqref{cH} and the spin dynamical model~\eqref{H}
are related via a slight modification of the usual freezing trick mechanism. We shall
exploit this connection to compute in closed form some eigenstates of $\cH$ with integer energy,
for any number of particles and arbitrary spin. We shall then compare our
results with those obtained by numerical diagonalization of $\cH$ for small values of
$N$ and $n$. The numerical computations strongly suggest that the exact states derived
in this section exhaust all the eigenstates of the chain~\eqref{cH} with integer energy.

We shall use in what follows the decomposition
\begin{equation}\label{Hsch}
H=\Hsc+2a\ms h(\bx)\,,
\end{equation}
where
\begin{equation}\label{h}
h(\bx)=\sum_i(x_i-x_{i+1})^{-2}\,(1-S_{i,i+1})\,.
\end{equation}
It shall also be convenient to write the scalar Hamiltonian~\eqref{Hsc} as
\begin{equation}\label{HscUV}
\Hsc=-\sum_i\pd_{x_i}^2+a^2U(\bx)-a\,V(\bx)\,,
\end{equation}
where the scalar potentials $U$ and $V$ are respectively given by
\begin{align}
&U(\bx)=r^2+\sum_i\frac2{(x_i-x_{i+1})^2}
+\sum_i\frac2{(x_i-x_{i-1})(x_i-x_{i+1})}\,,\label{U}\\[1mm]
&V(\bx)=\sum_i\frac{2}{(x_i-x_{i+1})^2}\,.\label{V}
\end{align}
{}From Eq.~\eqref{HscUV} it follows that for large $a$
the low-lying eigenfunctions of $\Hsc$
concentrate at absolute minima of the potential $U$ in the domain~\eqref{C} and their
energies satisfy~\cite{Si83,HK06}
\[
E=a^2\min_{\bx\in C} U(\bx)+O(a)\,.
\]
By comparison with the exact results for the ground state~\eqref{mu}, one concludes that
$\bxi$ is an absolute minimum of $U$ and $U(\bxi)=2N$. This condition has been used
in the previous section to check the accuracy of the numerical solution of the sites
equations~\eqref{sites} for large values of $N$.

There are two main limitations which prevent the use of the standard freezing
trick argument~\cite{Po94,EFGR05} for computing the spectrum of the spin
chain~\eqref{cH}. The first one is the requirement that $\bxi$ be the unique
minimum of the potential $U$ in the domain $C$. Although our numerical
calculations suggest that this is indeed the case, we have not been able to
provide a rigorous proof of this fact. The second limitation, which is more
fundamental, is the fact that the dynamical models $H$ and $\Hsc$ are only
quasi-exactly solvable. Let us briefly recall the basics of the usual freezing
trick method in order to understand why the full knowledge of the spectra of
$H$ and $\Hsc$ is essential for its application. Indeed, if the potential $U$
has an unique minimum $\bxi$ in $C$, for sufficiently large $a$ all the
eigenfunctions of $\Hsc$ are sharply peaked around this point. Thus, if
$\psi(\bx)$ is an eigenfunction of $\Hsc$ with energy $\Esc$ and $\ket{\si}$ is
an eigenstate of the chain $\cH$ with eigenvalue $\cE$, for $a\gg 1$ we have
\[
h(\bx)\psi(\bx)\ket{\si}\simeq\psi(\bx)h(\bxi)\ket{\si}\equiv\psi(\bx)\cH\ket{\si}
=\cE\psi(\bx)\ket{\si}\,.
\]
By Eq.~\eqref{Hsch}, the state $\psi(\bx)\ket{\si}$ is then an
approximate eigenfunction of $H$ with eigenvalue
\begin{equation}\label{Esc+cE}
E\simeq\Esc+2a\ms\cE\,.
\end{equation}
In other words, the Hamiltonian~\eqref{H} is approximately diagonal in a basis
of the form $\{\psi_i(\bx)\ket{\si_j}\}$, where $\{\psi_i(\bx)\}$ is a basis of
eigenfunctions of $\Hsc$ and $\{\ket{\si_j}\}$ is a basis of eigenstates of
$\cH$. Equation~\eqref{Esc+cE} cannot be used directly to compute the
corresponding spectrum of $\cH$, since it is not clear a priori which pairs of
eigenvalues of $H$ and $\Hsc$ yield an approximate eigenvalue $(E-\Esc)/(2a)$
of $\cH$. However, using Eq.~\eqref{Esc+cE} one can easily express the
partition function $\cZ$ of the spin chain $\cH$ in terms of the partition
functions $Z$ and $\Zsc$ of $H$ and $\Hsc$, respectively, via the formula
\begin{equation}\label{ZZZ}
\cZ(T)=\lim_{a\to\infty}\frac{Z(2aT)}{\Zsc(2aT)}\,.
\end{equation}
The latter formula, which is the key result behind the standard freezing trick
approach, cannot be used to compute the spectrum of the spin chain $\cH$
unless the whole spectrum of both $H$ and $\Hsc$ is known.

In spite of the above limitations, we shall see in this section that it is
still possible to compute a number of eigenstates of the spin chain~\eqref{cH}
from some of the families of spin eigenfunctions of the model~\eqref{H}
constructed in Ref.~\cite{EFGR07}. More precisely, we will show that certain
linear combinations of these eigenfunctions factorize as the product of the
ground state of the scalar model~\eqref{Hsc} times a spin function, whose
limit as $a\to\infty$ is an eigenstate of the spin chain Hamiltonian~$\cH$.

Let us begin by introducing some preliminary
notation. Let $\Si$ be the space of internal degrees of freedom of $N$ particles
with $\mathrm{su}(n)$ spin, and denote the elements of
its canonical basis by $\ket{s_1\dots s_N}$, where $s_i=-M,-M+1,\dots,M$ and $M=(n-1)/2$.
Let $\La$ be the total symmetrizer under particle permutations, that is
\[
\La=\frac1{N!}\,\sum_{k=1}^{N!}\Pi_k\,,
\]
where $\Pi_k$ denotes a permutation operator acting simultaneously
on spatial coordinates and spins. We shall consider in what follows the subspace
$\Si'\subset\Si$ of spin states $\ket{s}$ such that
$\sum_i\ket{s_{i,i+1}}$ is symmetric, where $\ket{s_{ij}}$ is defined by
\begin{equation}
\La\big(x_1x_2\ket s\big)=\sum_{i<j}x_ix_j\ket{s_{ij}}\,;\label{Lasij}\\
\end{equation}
see Ref.~\cite{EFGR07} for a complete characterization of this subspace.
Finally, given a state $\ket s\in\Si$ we define the spin functions
\begin{align}
&\Phi^{(k)}(\bx\ms;\!\ket s)=\La(x_1^k\ket s)\quad (k=0,1,2)\,,\qquad\tPhi^{(2)}(\bx\ms;\!\ket s)=\La(x_1x_2\ket s)\,,
\label{Phis}\\[1mm]
&\hPhi^{(3)}(\bx\ms;\!\ket s)=\La(x_1x_2(x_1-x_2)\ket s)\,.\label{hPhi}
\end{align}
We shall suppress one or both of the arguments of the above spin functions
when appropriate.

We shall next present the basic result we have used to construct eigenstates
of the chain~\eqref{cH} out of the eigenfunctions of the dynamical spin
model~\eqref{H} in Appendix~\ref{appA}. Recall that the energies of these
eigenfunctions are the numbers $E_{lm}=E_0+2a(2l+m)$, where $l$ and $m$ are
non-negative integers and $E_0$ is the ground state energy of the scalar
model~$\Hsc$.

The key point in the ensuing argument is the fact that for large $a$ the
normalized ground state $\mu_0=\mu/\|\mu\|$ of $\Hsc$, where $\mu$ is given
by~\eqref{mu} and $\|\mu\|^2=\int_C \mu^2$, is sharply peaked around its maximum
$\bxi$. Hence, if $F(\bx)$ is a continuous spin-dependent function such that
the integral of $\mu_0^2F$ over $C$ is finite, the main contribution to this
integral comes from a small ball centered at $\bxi$, up to exponentially small
terms. By the standard argument behind the proof of Laplace's method, it
follows that
\begin{equation}\label{mu0delta}
\lim_{a\to\infty}\int_C\mu_0^2F=F(\bxi)\lim_{a\to\infty}\int_C\mu_0^2=F(\bxi)\,.
\end{equation}

Assume now that $\Psi(\bx;a)=\mu_0(\bx;a)\Phi(\bx;a)$ is a continuous
eigenfunction of $H$ with energy $E_{kl}$ such that
$\ds\Phi(\bx;a)=\sum_{i=0}^ka^{-i}\Phi_i(\bx)$. Denoting by
\[
(f,F)=\int_C \overline{f(\bx)}F(\bx)\d^N\bx
\]
the usual spin-valued inner product of a scalar function $f(\bx)$
with a spin-valued function $F(\bx)$, from Eq.~\eqref{Hsch} we
obtain
\begin{equation}\label{Ekl}
E_{kl}(\mu_0^2,\Phi)= E_{kl}(\mu_0,\Psi)=
(\mu_0,H\Psi)=(\mu_0,\Hsc\Psi)+2a(\mu_0,h\Psi)\,.
\end{equation}
Since $\Hsc$ is self-adjoint and $h$ is a matrix multiplication operator,
the RHS of the previous equation equals
\[
(\Hsc\mu_0,\mu_0\Phi)+2a(\mu_0^2,h\Phi)=E_0(\mu_0,\mu_0\Phi)+2a(\mu_0^2,h\Phi)
=\big(\mu_0^2,(E_0+2ah)\Phi\big)\,.
\]
{}From Eq.~\eqref{Ekl} it follows that
\[
0=\big(\mu_0^2,(h-2l-m)\Phi\big)=\sum_{i=0}^ka^{-i}\big(\mu_0^2,(h-2l-m)\Phi_i\big)
\]
Taking the limit $a\to\infty$, and using
Eq.~\eqref{mu0delta} and the fact that $\cH=h(\bxi)$, we
finally obtain
\begin{equation}\label{cHsi}
(\cH-2l-m)\Phi_0(\bxi)=0\,.
\end{equation}
Thus
\begin{equation}\label{si}
\Phi_0(\bxi)=\lim_{a\to\infty}\Phi(\bxi)
\end{equation}
is either zero or an eigenstate of the
chain~\eqref{cH} with integer energy $2l+m$.

We shall now see that the application of the previous method to the
eigenfunctions~\eqref{Psi0}--\eqref{hPsi3} of the dynamical model~\eqref{H} listed in
Appendix~\ref{appA} yields the following types of eigenstates of the
chain~\eqref{cH}:
\begin{subequations}\label{s}
\begin{align}
&\vp_0(\ket s)=\Phi^{(0)}(\ket s)\equiv\La\ket s\,,\label{s0}\\[.5mm]
&\vp_1(\ket s)=\Phi^{(1)}(\bxi\ms;\!\ket s)\,,\label{s1}\\[.5mm]
&\vp_2(\ket s)=\Phi^{(2)}(\bxi\ms;\!\ket s)+(N-1)\tPhi^{(2)}(\bxi\ms;\!\ket s)\,,\label{s2}\\[.5mm]
&\vp_3(\ket s)=\hPhi^{(3)}(\bxi\ms;\!\ket s)+2\Phi^{(1)}(\bxi\ms;\!\ket s)\,.\label{s3}
\end{align}
\end{subequations}
For the states~\eqref{s2} $\ket s$ must belong to the subspace
$\Si'$ and be symmetric in the first two spins, whereas for the
states of type~\eqref{s3} $\ket s$ should be antisymmetric under
$S_{12}$. In all cases, the energy of the states $\vp_k(\ket s)$ is the integer $k$. We shall see in what follows that the states~\eqref{s3}
of energy $3$ only appear for spin $M>1/2$ ($n>2$).

Note, first of all, that Eq.~\eqref{La} implies that $\bxi^2=N$, since otherwise
each eigenfunction $\Psi^{(0)}_{l0}$ (cf.~Eq.~\eqref{Psi0}) would
yield an eigenstate of the chain~\eqref{cH} with eigenvalue $2l$ for arbitrary $l$.
(The fact that $\bxi^2=N$ can also be established directly from Eq.~\eqref{sites}
using the identity $U(\bxi)=2N$ for the scalar potential~\eqref{U}.)
Hence we need only consider the eigenfunctions~\eqref{Psi0}--\eqref{hPsi3} with $l=0$.
Moreover, from Eqs.~\eqref{Psi0}--\eqref{hPsi3} and~\eqref{BxP}
it follows that the eigenfunctions $\Psi^{(0)}_{0m}$,
$\Psi^{(2)}_{0m}$ and $\tPsi^{(2)}_{0m}$ (respectively $\Psi^{(1)}_{0m}$ and~$\hPsi^{(3)}_{0m}$)
vanish identically at $\bx=\bxi$ for odd (respectively even) $m$.

For the first type of eigenfunctions~\eqref{Psi0}, when $m$ is even Eq.~\eqref{Pa} implies that
\[
\Phi^{(0)}_{0m}(\bx)\equiv a^{-\frac m2}\mu_0^{-1}\Psi^{(0)}_{0m}\underset{a\to\infty}{\longrightarrow}
\frac{N^{\frac m2}}{\big(\frac m2\big)!}\,\Bx^m\,\Phi^{(0)}\,,
\]
where $\Phi^{(0)}$ is defined in~\eqref{Phis}. By Eq.~\eqref{xi=0}, the RHS of
the previous equation vanishes at $\bx=\bxi$ unless $m=0$. Thus the
eigenfunctions~\eqref{Psi0} only yield the zero energy eigenstates~\eqref{s0}
of $\cH$ (when $l=m=0$). Note that the fact that any symmetric spin state is
an eigenstate of the chain $\cH$ with zero energy follows directly from
Eq.~\eqref{cH}. Conversely, the latter equation implies that the ground state
energy of $\cH$ is zero and that the corresponding eigenstates are totally
symmetric. For this reason we shall concentrate in what follows on the
nontrivial states~\eqref{s1}--\eqref{s3}.

Let us next examine the eigenfunctions of type~\eqref{Psi1} for $l=0$ and odd $m\geq1$.
In this case we have
\begin{equation}\label{Phi1a}
\Phi^{(1)}_{0m}(\bx)\equiv a^{\frac{1-m}2}\mu_0^{-1}\Psi^{(1)}_{0m}\underset{a\to\infty}{\longrightarrow}
\frac{N^{\frac{m-1}2}}{\big(\frac{m-1}2\big)!}\,\Bx^{m-1}\big(\Phi^{(1)}-\Bx\ms\Phi^{(0)}\big)\,,
\end{equation}
where $\Phi^{(1)}$ is given by~\eqref{Phis}. Using again Eq.~\eqref{xi=0}
we conclude that $\lim_{a\to\infty}\Phi^{(1)}_{0m}$ vanishes at $\bx=\bxi$ unless $m=1$.
In this case, we have
\[
\lim_{a\to\infty}\Phi^{(1)}_{01}(\bxi)=\Phi^{(1)}(\bxi\ms;\!\ket s)\equiv\vp_1(\ket s)\,.
\]

Consider now the eigenfunctions $\Psi^{(2)}_{0m}$ and $\tPsi^{(2)}_{0m}$ with even $m$.
Using again Eq.~\eqref{Pa} we have
\begin{subequations}\label{Psis2a}
\begin{align}
a^{-\frac m2}\mu_0^{-1}\Psi^{(2)}_{0m}&=
-\frac{2N^{\frac m2}\,\Bx^m}{(m-1)\big(\tfrac{m-2}2\big)!}\,\Phi^{(0)}+O(a^{-1})\,,\label{Psi2a}\\[1mm]
a^{-\frac m2}\mu_0^{-1}\tPsi^{(2)}_{0m}&=\frac{2N^{\frac m2}\,\Bx^m}{(N-1)(m-1)\big(\tfrac{m-2}2\big)!}\,
\Phi^{(0)}+O(a^{-1})\,,\label{tPsi2a}
\end{align}
\end{subequations}
where $m\geq 2$. Thus the $O(1)$ term of the left-hand sides of Eqs.~\eqref{Psis2a} vanish
at $\bx=\bxi$ for all $m\geq 2$, on account of Eq.~\eqref{xi=0}. However,
if $\Psi^{(2)}_{0m}$ and $\tPsi^{(2)}_{0m}$ are built from the same spin state
$\ket s$, the previous equations imply that
the $O(1)$ part of the linear combination
$a^{-\frac m2}\mu_0^{-1}\big(\Psi^{(2)}_{0m}+(N-1)\tPsi^{(2)}_{0m}\big)$
vanishes. This observation suggests considering the function
\begin{align*}
\Phi^{(2)}_{0m}(\bx)&\equiv a^{1-\frac m2}\mu_0^{-1}\big(\Psi^{(2)}_{0m}+(N-1)\tPsi^{(2)}_{0m}\big)\\[1mm]
&=\frac{N^{\frac m2-1}}{\big(\tfrac{m-2}2\big)!}\,\Bx^{m-2}\big[\Phi^{(2)}+(N-1)\tPhi^{(2)}
-N\Bx\big(2\Phi^{(1)}-\Bx\ms\Phi^{(0)}\big)\big]+O(a^{-1})\,,
\end{align*}
where $m\geq 2$ is even and $\Phi^{(2)}$, $\tPhi^{(2)}$ are given in Eq.~\eqref{Phis}.
Note that $\Phi^{(2)}_{0m}$ is only defined
for states $\ket s\in\Si'$ symmetric under $S_{12}$, since otherwise $\tPsi^{(2)}_{0m}$
would not be defined.
The limit of $\Phi^{(2)}_{0m}$ as $a\to\infty$ vanishes at $\bx=\bxi$ unless $m=2$,
in which case we obtain the eigenstate with energy $2$
\[
\lim_{a\to\infty}\Phi^{(2)}_{02}(\bxi)
=\Phi^{(2)}(\bxi\ms;\!\ket s)+(N-1)\tPhi^{(2)}(\bxi\ms;\!\ket s)\equiv\vp_2(\ket s)\,.
\]

Let us finally turn to the last type of eigenfunctions~\eqref{hPsi3} with $l=0$ and odd $m\geq 3$.
{}From Eq.~\eqref{Pa} it immediately follows that
\begin{equation}\label{Psi3a}
a^{\frac{1-m}2}\mu_0^{-1}\hPsi^{(3)}_{0m}=
\frac{4N^{\frac{m-1}2}\,\Bx^{m-1}}{(m-2)\big(\frac{m-3}2\big)!}
\,\Big(\Phi^{(1)}-\frac{\Bx}{m}\,\Phi^{(0)}\Big)+O(a^{-1})\,,
\end{equation}
whose limit as $a\to\infty$ vanishes identically at $\bx=\bxi$ since $m\geq 3$ in this case.
However, as for the previous states, one can cancel the leading term in $a$
of $\hPsi^{(3)}_{0m}$ with a suitable linear combination
of $\Psi^{(0)}_{0m}$ and $\Psi^{(1)}_{0m}$ all built from the same spin state
(necessarily antisymmetric under $S_{12}$ for $\hPsi^{(3)}_{0m}$ to be defined).
Indeed, using Eqs.~\eqref{Phi1a} and~\eqref{Psi3a}
and taking into account that for odd $m$
\[
a^{\frac{1-m}2}\mu_0^{-1}\Psi^{(0)}_{0m}=
\frac{N^{\frac{m-1}2}}{\big(\frac{m-1}2\big)!}\,\Bx^m\,\Phi^{(0)}+O(a^{-1})\,,
\]
one can easily check that
\begin{equation}
\Phi^{(3)}_{0m}(\bx)\equiv a^{\frac{3-m}2}\mu_0^{-1}\Big(\hPsi^{(3)}_{0m}-\frac{2(m-1)}{m-2}\,\Psi^{(1)}_{0m}
-\frac{2(m-1)^2}{m(m-2)}\,\Psi^{(0)}_{0m}\Big)=O(1)\,.
\end{equation}
We shall next see that the $O(1)$ term of $\Phi^{(3)}_{0m}(\bx)$ vanishes at $\bx=\bxi$ unless $m=3$.
Indeed, Eq.~\eqref{BxP} implies that for $k'>2k$ the term $\Bx^{k'}P^{(\al+i,\be)}_k(t)$
vanishes at $\bx=\bxi$ to all orders in $a$. Hence the $O(1)$ terms of $\Phi^{(3)}_{0m}(\bx)$
and
\begin{equation}\label{O1}
a^{\frac{3-m}2}\Bx^{m-3}
\Big(P^{(\al+3,\be)}_{\frac{m-3}2}(t)\,\hPhi^{(3)}
-2\frac{m-1}{m-2}\,\Bx^2P^{(\al+1,\be)}_{\frac{m-1}2}(t)\,\Phi^{(1)}\Big)
\end{equation}
coincide at $\bx=\bxi$. {}From Eq.~\eqref{BxP} it is straightforward to show that the $O(1)$
term of~\eqref{O1} is a linear combination of $\Bx^{m-3}$ and $\Bx^{m-1}$, thus establishing
our claim.

By the previous remarks, we need only compute the $O(1)$ term of~\eqref{O1} for $m=3$, which
by Eq.~\eqref{Pa} is given by
\[
\hPhi^{(3)}+\frac2N\,\big(r^2-N(N+2)\Bx^2\big)\Phi^{(1)}\,.
\]
Using Eq.~\eqref{xi=0} and the identity $\bxi^2=N$ we finally obtain
the eigenstates~\eqref{s3} of energy $3$:
\[
\lim_{a\to\infty}\Phi^{(3)}_{03}(\bxi)
=\hPhi^{(3)}(\bxi\ms;\!\ket s)+2\ms\Phi^{(1)}(\bxi\ms;\!\ket s)\equiv\vp_3(\ket s)\,.
\]

The previous discussion guarantees that the nontrivial states $\vp_k(\ket s)$
given by Eqs.~\eqref{s1}--\eqref{s3} are eigenstates of the chain $\cH$ with energy
$k$ provided that they do not vanish. For instance, the states~\eqref{s1} are easily seen
to vanish when the spin state $\ket s$ is symmetric, since in this case
\[
\vp_1(\ket s)=\Phi^{(1)}(\bxi\ms;\!\ket s)=\Bxi\ket s=0\,.
\]
This is also the case for the states~\eqref{s2}. Indeed, if $\ket s$ is symmetric
it clearly belongs to $\Si'$ and we have
\[
\Phi^{(2)}(\bx\ms;\!\ket s)+(N-1)\tPhi^{(2)}(\bx\ms;\!\ket s)
=\frac{r^2}N\,\ket s+\frac2N\,\sum_{i<j}x_ix_j\,\ket s
=N\Bx^2\ket s\,.
\]
Since $\vp_2(\ket s)$ coincides with the LHS of this expression evaluated at $\bx=\bxi$,
it vanishes on account of Eq.~\eqref{xi=0}.
Less trivially, let $\ket s$ be a linear combination of basic states
$\ket{\bs}\equiv\ket{s_1\ldots s_N}$ such that $\vert\bs\vert\leq 2$,
where $\vert\bs\vert$ denotes the number of distinct components of
$\bs\equiv(s_1,\dots,s_N)$. If, in addition, $\ket s$ is antisymmetric under $S_{12}$,
we shall prove below that
\begin{equation}\label{hPhi3n2}
\hPhi^{(3)}+2\Big(\frac{r^2}N\,\Phi^{(1)}-\Bx\Phi^{(2)}\Big)=0\,,
\end{equation}
which implies that $\vp_3(\ket s)=0$ by Eq.~\eqref{xi=0}
and the identity $\bxi^2=N$. In summary, the state $\ket s$ in the definition
of the states~\eqref{s1}--\eqref{s3} can be taken without loss of generality as follows:
\begin{align}
&\vp_1(\ket s):\qquad\qquad\ket s\in\Si-\La(\Si)\,;\label{c1}\\[1mm]
&\vp_2(\ket s):\qquad\qquad \ket s\in\Si'-\La(\Si)\,,\quad S_{12}\ket s=\ket s\,;\label{c2}\\[1mm]
&\vp_3(\ket s):\qquad\qquad \ket s=\sum_{|\bs|\geq3}c_\bs\ket\bs\,,\quad S_{12}\ket s=-\ket s\,.
\label{c3}
\end{align}
Note, in particular, that the states~\eqref{s3} do not appear for spin $1/2$ in view of
the first condition~\eqref{c3}.

Our next task is to study the number of linearly independent
states of the form~\eqref{s1}--\eqref{s3} with $\ket s$ satisfying the above conditions.
The main difficulty in this respect is the fact that the eigenstates $\vp_k(\ket s)$
constructed from a set of linearly independent states $\ket s$
satisfying conditions~\eqref{c1}--\eqref{c3} need not be
independent. In order to address this problem, it is convenient to introduce
some additional notation. Given a basic state $\ket\bs$ with
$\vert\bs\vert=p$, we define its {\em spin content} as the set of pairs $(s^i,\nu_i)$,
where $s^1<\dots<s^p$ are the
distinct components of $\bs$ and $\nu_i$ is the number of times that $s^i$ appears in $\bs$.
For instance, the spin content of the basic state
$\ket\bs=\ket{0,{-2},1,-2,1}$ is $\{(-2,2),(0,1),(1,2)\}$.
We shall say that an arbitrary spin state $\ket s$ has well-defined spin content
if it is a linear combination of basic states having the same spin content.
It is obvious that a set of states whose spin contents are all different is linearly independent.
Note also that if $\ket s$ has a well-defined spin content, then $\vp_k(\ket s)$
has the same spin content. Therefore, it suffices to determine the number of
linearly independent states $\vp_k(\ket s)$ built from states $\ket s$ with well-defined
spin content. Clearly, for a given spin content $\{(s^i,\nu_i)\}$ this number is independent
of the particular values of the spin coordinates $s^i$. We shall therefore denote by
$d_k(\nu_1,\dots,\nu_p)\equiv d_k(\bnu)$ the dimension of the linear space of states $\vp_k(\ket s)$ with
a given spin content $\{(s^i,\nu_i)\}$. We shall next prove the following upper bounds
on these dimensions:
\begin{equation}\label{dks}
d_1(\bnu)\leq p-1\,,\qquad d_2(\bnu)\leq p-1\,,\qquad d_3(\bnu)\leq\binom{p-1}{2}\,.
\end{equation}

Consider in the first place the states of the form $\vp_1(\ket s)$
with a given spin content. If $\ket\bs$ and $\ket{\bs'}$ are two
basic states with the same spin content and differing by a
permutation of the last $N-1$ spin coordinates, then
$\La(x_1\ket\bs)=\La(x_1\ket{\bs'})$, which implies that
$\vp_1(\ket\bs)=\vp_1(\ket{\bs'})$ by Eqs.~\eqref{Phis}
and~\eqref{s1}. Hence, the space of states $\vp_1(\ket s)$ with
spin content $\{(s^1,\nu_1),\dots,(s^p,\nu_p)\}$ is spanned by the
states
\begin{equation}\label{vp1}
\vp_{1}(\ket{s^i\dots})\,,\qquad i=1,\dots,p\,,
\end{equation}
where the ellipsis denotes any ordering of the remaining $N-1$
spin components corresponding to the above spin content. On the
other hand, the $p$ states~\eqref{vp1} satisfy the linear relation
\begin{equation}\label{sumvp1}
\sum_{i=1}^p\nu_i\,\vp_{1}(\ket{s^i\dots})=0\,,
\end{equation}
which implies the first inequality in~\eqref{dks}. Indeed, if
$\ket\bs$ is a basic state with spin content $\{(s^i,\nu_i)\}$, we
have
\[
N\Bx\La\ket\bs=\sum_{i=1}^N\La\big(x_i\ket\bs\big)
=\sum_{i=1}^p\nu_i\,\La\big(x_1\ket{s^i\dots}\big)\,.
\]
Setting $\bx=\bxi$ and using~\eqref{xi=0} we immediately
obtain~\eqref{sumvp1}.

Let us turn now to the states of the form $\vp_2(\ket s)$, where $\ket s$ satisfies~\eqref{c2}
and has a well-defined spin content $\{(s^1,\nu_1),\dots,(s^p,\nu_p)\}$.
The results of our previous paper~\cite{EFGR07} (see Proposition~4)
imply that the space of states of this form is spanned
by
\begin{equation}\label{vp2}
\vp_2(\ket{\chi_i})\,,\qquad i=1,\dots,p\,,
\end{equation}
where
\begin{equation}\label{chis}
\ket{\chi_i}=
\begin{cases}
\ds\nu_i(\nu_i-1)\ket{s^is^i\dots}-\sum_{\substack{1\leq j,k\leq p\\j,k\neq i}}\nu_j(\nu_k-\de_{jk})
\ket{s^js^k\dots}\,,\quad &\nu_i>1\,,\\
\ds\sum_{\substack{1\leq j\leq p\\j\neq i}}\nu_j\,
\big(\ket{s^is^j\dots}+\ket{s^js^i\dots}\big)\,,\quad &\nu_i=1\,.\vrule height20pt width0pt
\end{cases}
\end{equation}
However, from \cite[Prop.~3]{EFGR07} it follows that the $p$ states~\eqref{vp2}
satisfy the linear relation
\begin{equation}\label{sumvp2}
\sum_{i=1}^p\vp_2(\ket{\chi_i})=0\,,
\end{equation}
which yields the second inequality in~\eqref{dks}.

Consider finally the states of the form $\vp_3(\ket s)$, where $\ket s$
is antisymmetric under $S_{12}$ and has a certain spin content
$\{(s^1,\nu_1),\dots,(s^p,\nu_p)\}$. Clearly, the space of such states
is spanned by
\begin{equation}\label{vp3}
\vp_3\big(\ket{s^is^j\dots}-\ket{s^js^i\dots}\big)\,,\qquad 1\leq i<j\leq p\,.
\end{equation}
We shall prove below that the latter $\binom p2$ states satisfy
the linear relations
\begin{equation}\label{rels}
\sum_{j=1}^p\nu_j\,\vp_3\big(\ket{s^is^j\dots}-\ket{s^js^i\dots}\big)=0\,,
\qquad i=1,\dots,p\,.
\end{equation}
Multiplying the LHS of the $i$-th relation by $\nu_i$ and summing over $i$ we obtain
\[
\sum_{i,j=1}^p\nu_i\nu_j\,\vp_3\big(\ket{s^is^j\dots}-\ket{s^js^i\dots}\big)\,,
\]
which vanishes by antisymmetry. Thus we can safely drop one of the identities~\eqref{rels},
say the last one, and the first $p-1$ are independent since they can be solved
for $\vp_3\big(\ket{s^is^p\dots}-\ket{s^ps^i\dots}\big)$ for $i=1,\dots,p-1$.
Hence there are at most $\binom p2-(p-1)=\binom{p-1}2$ independent states of the form~\eqref{vp3},
which establishes the last inequality in~\eqref{dks}.

We still have to prove the identities~\eqref{rels}. To this end,
denote by $\Phi_3\equiv\Phi_3(\bx\ms;\!\ket s)$ the LHS of Eq.~\eqref{hPhi3n2}.
Since $\Phi_3(\bxi\ms;\!\ket s)=\vp_3(\ket s)$, Eqs.~\eqref{rels}
follow directly from the identities
\begin{equation}\label{Phi3rels}
\sum_{j=1}^p\nu_j\,
\Phi_3\big(\bx\ms;\!\ket{s^is^j\dots}-\ket{s^js^i\dots}\big)=0\,,
\qquad i=1,\dots,p\,,
\end{equation}
In order to establish~\eqref{Phi3rels}, note first that
\begin{align*}
r^2\Phi^{(1)}\big(\ket{s^is^j\ldots}\big)-N\Bx\ms\Phi^{(2)}\big(\ket{s^is^j\ldots}\big)
&=-\sum_{l=1}^N\La\big(x_1x_l(x_1-x_l)\ket{s^is^j\ldots}\big)\notag\\
&=-\sum_{k=1}^p\nu_k\hPhi^{(3)}\big(\ket{s^is^k\ldots}\big)\,,
\end{align*}
so that
\begin{multline}\label{sumPhi3}
\frac N2\sum_{j=1}^p\nu_j\,
\Phi_3\big(\ket{s^is^j\dots}-\ket{s^js^i\dots}\big)
=N\sum_{j=1}^p\nu_j\,
\hPhi^{(3)}\big(\ket{s^is^j\dots}\big)\\
-\sum_{j,k=1}^p\nu_j\nu_k\,
\hPhi^{(3)}\big(\ket{s^is^k\dots}\big)+\sum_{j,k=1}^p\nu_j\nu_k\,
\hPhi^{(3)}\big(\ket{s^js^k\dots}\big)\,.
\end{multline}
Since $\sum_{j=1}^p\nu_j=N$, the first two terms of the RHS of the previous formula
cancel, whereas the last one vanishes by antisymmetry. This concludes the proof
of Eq.~\eqref{Phi3rels}. Note, finally, that the latter equation for $p=2$
easily yields the identity~\eqref{hPhi3n2}.

\subsection{Example} We shall now discuss in detail the case $N=5$ and spin
$M=1$. The sites of this chain are given by $\xi_5=-\xi_1=1.469$,
$\xi_4=-\xi_2=0.584$, $\xi_3=0$. The Hamiltonian~\eqref{cH} is
represented by a $243\times 243$ real symmetric matrix, whose
numerical diagonalization is straightforward. The spectrum
consists of $21$ different levels, the only integer energies being
$0,1,2,3$. For conciseness' sake, we only present the first six
levels and the last one with their corresponding degeneracies,
cf.~Table~\ref{table:example}. Notice, in particular, the
appearance of a noninteger energy between the third and the fifth
levels. The degeneracy of the ground level is given by the number
of independent symmetric states, that is, the number of
combinations with repetitions of $5$ elements from $3$. Turning to
the other integer energies, we shall next present a basis of
eigenstates of each nontrivial type~\eqref{s1}--\eqref{s3}. By the
previous discussion, we must first determine all the possible spin
contents $\{(s^i,\nu_i)\}$ compatible with
conditions~\eqref{c1}--\eqref{c3}, and then construct a basis of
states $\vp_k(\ket s)$ with each of these spin contents. In
practice, the former task is performed in two steps, namely, one
first finds all possible degeneracy vectors $\bnu$, and then one
determines the spin contents associated with each of these
vectors.

\begin{table}[h]
\caption{Energies and degeneracies of the first six and the last
levels of the chain~\eqref{cH} with $N=5$ and
$M=1$.\vskip2mm}\label{table:example}
\begin{center}
\begin{tabular}{c@{\hspace*{2.5em}}c}\hline
  \vrule height 11pt depth 7pt width0pt Energy&
  Degeneracy\\\hline
0& 21\\ \hline
1& 24\\ \hline
2& 24\\ \hline
2.035& 15\\ \hline
3& 6\\ \hline
3.953& 15\\ \hline
$\cdots$& $\cdots$\\\hline
15.033& 3\\\hline
\end{tabular}
\end{center}
\end{table}

The states~\eqref{s1} of unit energy are generated by basic states
$\ket\bs$ with $p\geq 2$ different spin components. In
Table~\ref{table:vp1} we present the list of such basic states,
where we have taken into account the relation~\eqref{sumvp1} to
drop one state for each different spin content; for instance,
the state $\ket{0\ms{-}{-}{-}{-}}$ does not appear in the table,
since $\vp_1(\ket{\ms
0\ms{-}{-}{-}{-}})=-4\vp_1(\ket{{-}{-}{-}{-}\ms 0})$. In this way
we obtain~$24$ basic states, whose corresponding states
$\vp_1(\ket\bs)$ turn out to be linearly independent.

\begin{table}[h]
\caption{List of basic states $\ket\bs$ generating the
states~\eqref{s1} of unit energy for each degeneracy vector
$\bnu$.}\vskip2mm\label{table:vp1}
\setlength\tabcolsep{2pt}
\setbox0=\hbox{$\ket{{-}{-}{-}{-}{+}}$,}
\begin{center}
\begin{tabular}{@{\hspace*{.5em}}l@{\hspace*{2.5em}}lll}\hline
  \vrule height 11pt depth 7pt width0pt \quad$\bnu$ &
  & \parbox{\wd0}{\centering$\ket\bs$}&\\ \hline
$(4,1)$ & $\ket{{-}{-}{-}{-}\mms0}$, & $\ket{{-}{-}{-}{-}{+}}$, & $\ket{\mms0\mms0\mms0\mms0{+}}$\\ \hline
$(3,2)$ & $\ket{{-}{-}{-}\mms0\mms0}$, & $\ket{{-}{-}{-}{+}{+}}$,& $\ket{\mms0\mms0\mms0{+}{+}}$\\ \hline
$(2,3)$ & $\ket{{-}{-}\mms0\mms0\mms0}$, & $\ket{{-}{-}{+}{+}{+}}$, & $\ket{\mms0\mms0{+}{+}{+}}$\\ \hline
$(1,4)$ & $\ket{{-}\mms0\mms0\mms0\mms0}$, & $\ket{{-}{+}{+}{+}{+}}$, & $\ket{\mms0{+}{+}{+}{+}}$\\ \hline
$(3,1,1)$ & $\ket{{-}{-}{-}\mms0{+}}$, & $\ket{\mms0{-}{-}{-}{+}}$
&\\ \hline
$(2,2,1)$ & $\ket{{-}{-}\mms0\mms0{+}}$, & $\ket{\mms0\mms0{-}{-}{+}}$ &\\ \hline
$(2,1,2)$ & $\ket{{-}{-}\mms0{+}{+}}$, & $\ket{\mms0{-}{-}{+}{+}}$ &\\ \hline
$(1,3,1)$ & $\ket{{-}\mms0\mms0\mms0{+}}$, & $\ket{\mms0\mms0\mms0{-}{+}}$ &\\ \hline
$(1,2,2)$ & $\ket{{-}\mms0\mms0{+}{+}}$, & $\ket{\mms0\mms0{-}{+}{+}}$ &\\ \hline
$(1,1,3)$ & $\ket{{-}\mms0{+}{+}{+}}$, & $\ket{\mms0{-}{+}{+}{+}}$ &\\ \hline
\end{tabular}
\end{center}
\end{table}

Similarly, the space of states $\vp_2(\ket s)$ of energy $2$ is spanned by the states~\eqref{vp2}
with $\ket{\chi_i}$ given by~\eqref{chis}, for each spin content
$\{(s^1,\nu_1),\dots,(s^p,\nu_p)\}$
such that $p\geq 2$. As in the previous case, the relation~\eqref{sumvp2}
implies that for each spin content one can drop one of the states $\ket{\chi_i}$.
In Table~\ref{table:vp2} we present a possible choice of the $24$ states $\ket{\chi_i}$
that can be constructed by this procedure. Just as before, the corresponding states
$\vp_2(\ket{\chi_i})$ are easily seen to be linearly independent.

\begin{table}[h]
\caption{States $\ket\chi$ of the form~\eqref{chis} generating the
states~\eqref{vp2} of energy $2$ for each degeneracy vector
$\bnu$.}\vskip2mm\label{table:vp2}
\setbox0=\hbox{$\ket{{-}\mms0{+}{+}{+}}+\ket{\mms0{-}{+}{+}{+}}
+3(\ket{{-}{+}{+}{+}\mms0}+\ket{{+}{-}\mms0{+}{+}})$\,,}
\begin{center}
\begin{tabular}{l@{\hspace*{2.5em}}l}\hline
\vrule height 11pt depth 7pt width0pt \quad$\bnu$& \parbox{\wd0}{\centering$\ket\chi$}\\ \hline
$(4,1)$ & $\ket{{-}{-}{-}{-}\mms0}$\,, $\ket{{-}{-}{-}{-}{+}}$\,, $\ket{\mms0\mms0\mms0\mms0{+}}$\\ \hline
$(3,2)$ & 3$\ket{{-}{-}{-}\mms0\mms0}-\ket{\mms0\mms0{-}{-}{-}}$\,,
$3\ket{{-}{-}{-}{+}{+}}-\ket{{+}{+}{-}{-}{-}}$\,,\\[-1.2mm]
& 3$\ket{\mms0\mms0\mms0{+}{+}}-\ket{{+}{+}\mms0\mms0\mms0}$\\ \hline
$(2,3)$ & $3\ket{\mms0\mms0\mms0{-}{-}}-\ket{{-}{-}\mms0\mms0\mms0}$\,,
$3\ket{{+}{+}{+}{-}{-}}-\ket{{-}{-}{+}{+}{+}}$\,,\\[-1.2mm]
& $3\ket{{+}{+}{+}\mms0\mms0}-\ket{\mms0\mms0{+}{+}{+}}$\\ \hline
$(1,4)$ & $\ket{\mms0\mms0\mms0\mms0{-}}$\,, $\ket{{+}{+}{+}{+}{-}}$\,, $\ket{{+}{+}{+}{+}\mms0}$\\ \hline
$(3,1,1)$ &
$6\ket{{-}{-}{-}\mms0{+}}-\ket{\mms0{+}{-}{-}{-}}-\ket{{+}\mms0{-}{-}{-}}$\,,\\[-1.2mm]
&$3(\ket{\mms0{-}{-}{-}{+}}+\ket{{-}\mms0{-}{-}{+}})
+\ket{\mms0{+}{-}{-}{-}}+\ket{{+}\mms0{-}{-}{-}}$\\ \hline
$(2,2,1)$ & $\ket{{-}{-}\mms0\mms0{+}}-\ket{\mms0\mms0{-}{-}{+}}
-\ket{\mms0{+}{-}{-}\mms0}-\ket{{+}\mms0\mms0{-}{-}}$\,,\\[-1.2mm]
& $\ket{\mms0\mms0{-}{-}{+}}-\ket{{-}{-}\mms0\mms0{+}}
-\ket{{-}{+}{-}\mms0\mms0}-\ket{{+}{-}{-}\mms0\mms0}$\\ \hline
$(2,1,2)$ & $\ket{{-}{-}\mms0{+}{+}}-\ket{{+}{+}{-}{-}\mms0}-\ket{\mms0{+}{+}{-}{-}}
-\ket{{+}\mms0{-}{-}{+}}$\,,\\[-1.2mm]
& $\ket{{+}{+}{-}{-}\mms0}-\ket{{-}{-}{+}{+}\mms0}
-\ket{{-}\mms0{-}{+}{+}}-\ket{\mms0{-}{-}{+}{+}}$\\ \hline
$(1,3,1)$ & $3(\ket{{-}\mms0\mms0\mms0{+}}+\ket{\mms0{-}\mms0\mms0{+}})
+\ket{{-}{+}\mms0\mms0\mms0}+\ket{{+}{-}\mms0\mms0\mms0}$\,,\\[-1.2mm]
& $6\ket{\mms0\mms0\mms0{-}{+}}-\ket{{-}{+}\mms0\mms0\mms0}-\ket{{+}{-}\mms0\mms0\mms0}$\\ \hline
$(1,2,2)$ & $\ket{\mms0\mms0{-}{+}{+}}-\ket{{+}{+}{-}\mms0\mms0}
-\ket{{-}{+}{+}\mms0\mms0}-\ket{{+}{-}\mms0\mms0{+}}$\,,\\[-1.2mm]
& $\ket{{+}{+}{-}\mms0\mms0}-\ket{\mms0\mms0{-}{+}{+}}
-\ket{{-}\mms0\mms0{+}{+}}-\ket{\mms0{-}\mms0{+}{+}}$\\ \hline
$(1,1,3)$ & $\ket{{-}\mms0{+}{+}{+}}+\ket{\mms0{-}{+}{+}{+}}
+3(\ket{{-}{+}{+}{+}\mms0}+\ket{{+}{-}\mms0{+}{+}})$\,,\\[-1.2mm]
& $6\ket{{+}{+}{+}{-}\mms0}-\ket{{-}\mms0{+}{+}{+}}-\ket{\mms0{-}{+}{+}{+}}$\\ \hline
\end{tabular}
\end{center}
\end{table}

Finally, for spin $1$ the set of states $\vp_3(\ket s)$ of energy $3$
is spanned by states of the form~\eqref{vp3} associated with spin
contents with three different components, cf.~Eq.~\eqref{c3}. By
the last inequality~\eqref{dks}, for each such spin content there
is at most one independent state of the form~\eqref{vp3}. Thus
there are at most $6$ independent states of the form~\eqref{s3},
generated (for instance) by the states $\ket s$ listed in
Table~\ref{table:vp3}. As in the previous cases, it can be checked
that the $6$ states $\vp_3(\ket s)$ constructed from the states in
Table~\ref{table:vp3} are actually independent.

\begin{table}[h]
\caption{List of states $\ket s$ generating the
states~\eqref{s3} of energy $3$ and their corresponding
degeneracy vectors $\bnu$.}\vskip2mm\label{table:vp3}
\setbox0=\hbox{$\ket{{-}\mms0{+}{+}{+}}-\ket{\mms0{-}{+}{+}{+}}$}
\begin{center}
\begin{tabular}{l@{\hspace*{2.5em}}l}\hline
\vrule height 11pt depth 7pt width0pt \quad$\bnu$
& \parbox{\wd0}{\centering$\ket s$}\\ \hline
$(3,1,1)$ & $\ket{\mms0{+}{-}{-}{-}}-\ket{{+}\mms0{-}{-}{-}}$\\ \hline
$(2,2,1)$ & $\ket{{+}{-}{-}\mms0\mms0}-\ket{{-}{+}{-}\mms0\mms0}$\\ \hline
$(2,1,2)$ & $\ket{\mms0{-}{-}{+}{+}}-\ket{{-}\mms0{-}{+}{+}}$\\ \hline
$(1,3,1)$ & $\ket{{-}{+}\mms0\mms0\mms0}-\ket{{+}{-}\mms0\mms0\mms0}$\\ \hline
$(1,2,2)$ & $\ket{{-}\mms0\mms0{+}{+}}-\ket{\mms0{-}\mms0{+}{+}}$\\ \hline
$(1,1,3)$ & $\ket{{-}\mms0{+}{+}{+}}-\ket{\mms0{-}{+}{+}{+}}$\\ \hline
\end{tabular}
\end{center}
\end{table}

Several important remarks can be made in connection with the
previous example. In the first place, it is apparent that the
inequalities~\eqref{dks} are in this case equalities for every
spin content. Secondly, the number of independent states with
integer energy $k$ of the form $\vp_k(\ket s)$ coincides with the
degeneracy of the corresponding level,
cf.~Table~\ref{table:example}. Hence, in this example {\em all}
the eigenstates with integer energy are of the form~\eqref{s}, and
can therefore be computed explicitly. In fact, we have performed a
similar study for the case of $N=6$ and spin $M=3/2$, arriving at
exactly the same conclusions. In view of these examples, it is
natural to formulate the following conjectures:

\begin{itemize}
\item[1.] The inequalities~\eqref{dks} are always equalities.\smallskip

\item[2.] The only integer energies of the chain~\eqref{cH} are $0,1,2$ and
(for $M\geq1$) $3$, corresponding to the first three and the fifth levels.\smallskip

\item[3.] The only eigenstates with integer energy are those of the form~\eqref{s}.
\end{itemize}

According to the first conjecture, for $k=1,2,3$ the number of independent states $\vp_k(\ket s)$
with a well-defined spin content with $p$ elements depends only on $p$, and is given by
\[
d_k(p)=%
\begin{cases}
\raise1mm\hbox{$p-1\,,$}\quad & \raise1mm\hbox{$k=1,2\,,$}\\[1mm]
\raise1mm\hbox{$\binom{p-1}2\,,$}\quad & \raise1mm\hbox{$k=3\,.$}
\end{cases}
\]
By the third conjecture, the degeneracy of the integer energy $k$ is thus
\begin{equation}\label{dk}
d_k=\sum_{p=1}^{\min(n,N)}d_k(p)\binom{n}{p}\binom{N-1}{p-1}\,,\qquad k=1,2,3\,.
\end{equation}
Indeed, there are $\binom np$ different choices of $p$ distinct spin values $s^i$,
and $\binom{N-1}{p-1}$ ways of selecting $p$ numbers $\nu_i\geq 1$ such that
$\nu_1+\dots+\nu_p=N$. In order to evaluate the sum in~\eqref{dk}, note first that
\begin{equation}\label{dksum}
\sum_{p=1}^{\min(n,N)}\binom{p-1}j\binom{n}{p}\binom{N-1}{p-1}
=\binom{N-1}j\sum_{p=1}^{\min(n,N)}\binom{n}{p}\binom{N-j-1}{p-j-1}\,.
\end{equation}
On the other hand, expanding both sides of the identity
\[
\big(1+z^{-1}\big)^n(1+z)^{N-j-1}=z^{-n}(1+z)^{N+n-j-1}
\]
in powers of $z$, we easily obtain
\[
\sum_{p=1}^{\min(n,N)}\binom{n}{p}\binom{N-j-1}{p-j-1}
=\binom{N+n-j-1}N\,.
\]
The previous identity and Eq.~\eqref{dksum} immediately yield
the following explicit formula for the degeneracy of the positive integer
energies of the chain~\eqref{cH}:
\begin{equation}\label{dkfinal}
d_k=\binom{N-1}{k-1+\de_{k1}}\binom{N+n-k-\de_{k1}}N\,,\qquad k=1,2,3\,.
\end{equation}
(The degeneracy of the ground level $\cE=0$ is the dimension of the space of
symmetric states, namely $\binom{N+n-1}{N}$\,.)

We have numerically diagonalized the chain Hamiltonian~\eqref{cH} for several
values of $n$ and $N$ such that $n^N\leq 3^8=6561$. In all cases, we have checked
the validity of the second conjecture, and that the formula~\eqref{dkfinal}
for the degeneracies of the integer energies (which is a direct
consequence of the first and third conjectures) is satisfied.
These results lend strong numerical support to the above conjectures,
whose rigorous proof deserves further investigation.

\section{Statistical analysis of the spectrum}\label{S:stas}

It has been recently shown that for large $N$ the level density of
spin chains of HS type can be approximated with remarkable
accuracy by a Gaussian distribution~\cite{FG05,EFGR05,BB06}. A
natural question is whether this is also the case for the spin
chain~\eqref{cH} under consideration. The fact that the partition
function of the chain~\eqref{cH} has not been computed in closed
form ---unlike those of the HS chains in the previous
references--- makes it difficult to address this problem directly.
Although one can in principle diagonalize numerically the matrix
of the chain Hamiltonian $\cH$, in practice this is only feasible
for relatively small values of $N$. In this section we shall study
the level density of the chain~\eqref{cH} using the
method of moments outlined in Appendix~\ref{appB}, which provides an
accurate estimation of the level density for larger values of $N$.

In order to improve the numerical stability of the method, it is convenient to
take the parameters $b$ and $c$ in Eq.~\eqref{om} as
\begin{equation}\label{bc}
b=0\,,\qquad c=1\,,
\end{equation}
so that the spectrum of $\cA$ lies in the interval $[-2,2]$.
Therefore the matrix $\cA$ and the Hamiltonian~$\cH$ are related by
\begin{equation}\label{cA}
\cA=\frac{4\cH}{\cE_{max}}-2\,,
\end{equation}
where $\cE_{\max}$ is the largest eigenvalue of~$\cH$. Note that
this eigenvalue can be computed numerically without
difficulty for relatively large values of $N$, since the matrix
of $\cH$ is very sparse. One must then check that the coefficients $b_k$ and
$c_k$ in the continued fraction expansion of the resolvent of $\cA$
(cf.~Appendix~\ref{appB}) approximately stabilize to the values $b$ and $c$ in
Eq.~\eqref{bc} for $k_0<k<k_1$, with $k_1\gg k_0$. We have verified that this
is indeed the case for $N$ sufficiently large, where $k_0$ is typically of the
order of $10$. We can thus use Eq.~\eqref{gfinal} to obtain a continuous
approximation $g(x)$, where $x\equiv (4\cE/\cE_{max})-2$, to the density of eigenvalues of the matrix $\cA$, as
explained in Appendix~\ref{appB}.

\begin{figure}[h]
\begin{center}
\psfrag{g}[Bc][Bc][1][0]{\footnotesize $g(x)$}
\psfrag{x}{\footnotesize $x$}
\includegraphics[height=6cm]{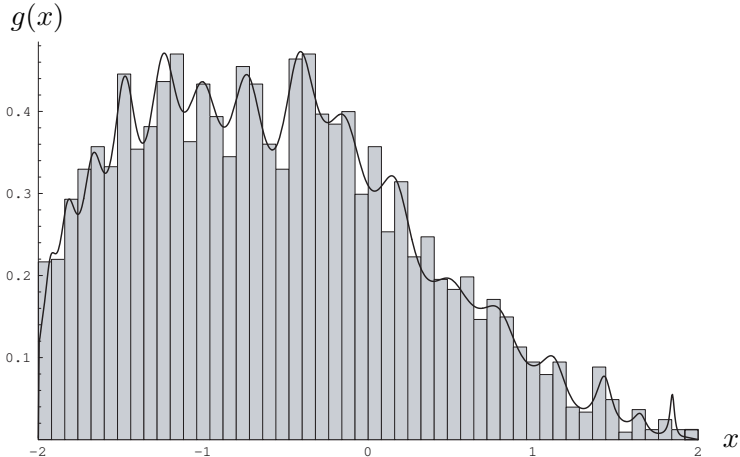}
\begin{quote}
  \caption{Continuous approximation $g(x)$ to the density of eigenvalues of
    the matrix~\eqref{cA} for spin $1/2$ and $N=12$
    ($\cE_{\max}=201.21$) compared with the histogram of the spectrum of this matrix.\label{fig:g12}}
\end{quote}
\end{center}
\end{figure}

As a test of the accuracy of the method, we have computed the
density $g$ for spin~$1/2$ and $N=12$, since in this case the
matrix~$\cA$ can still be diagonalized numerically. In
Fig.~\ref{fig:g12} we have compared the approximate density $g$
computed by applying the moments method with $k_0=20$ and $20$
random vectors with the histogram of the spectrum of $\cA$
obtained by subdividing the interval $[-2,2]$ in $50$
subintervals. As can be seen from the latter figure, the
continuous density $g$ essentially reproduces the shape of the
histogram.

\begin{figure}[h]
\begin{center}
\psfrag{g}[Bc][Bc][1][0]{\footnotesize $f(x),g(x)$}
\psfrag{x}{\footnotesize $x$}
\includegraphics[height=6cm]{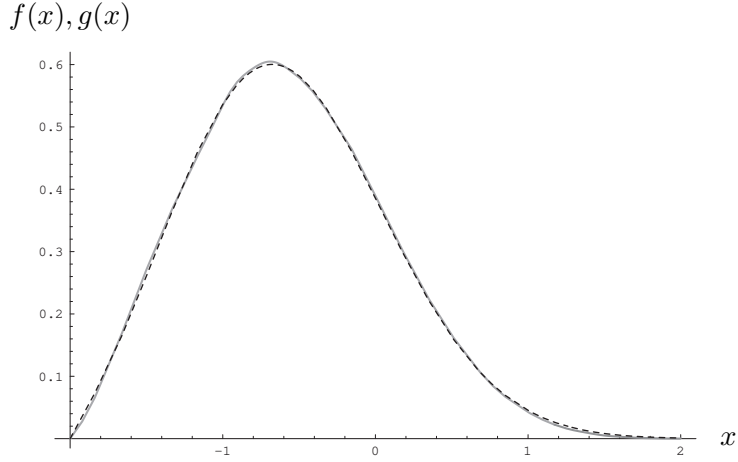}
\begin{quote}
\caption{Approximate density $g(x)$ of eigenvalues of
the matrix~\eqref{cA} (grey line) for spin $1/2$ and $N=21$ ($\cE_{\max}=1126.53$),
compared with the distribution \eqref{fdist} with optimal parameters
$\al=0.85$, $\be=1.27$, $\ga=0.91$ (dashed line).\label{fig:fg21}}
\end{quote}
\end{center}
\end{figure}

The real interest of the method of moments is the possibility of approximating
the density of eigenvalues of a large matrix whose numerical diagonalization
is not feasible. We have been able to compute the approximate density $g(x)$
of eigenvalues of the matrix $\cA$ for $N$ up to $21$ for spin $1/2$, and up to $13$ for
spin~$1$. For instance, in Fig.~\ref{fig:fg21} we present the plot of $g$ for spin
$1/2$ and $N=21$, computed with $k_0=20$ and $20$ random vectors. It is
apparent from this plot, and for similar plots for spins $1/2$ and $1$, that
the level density is not Gaussian. As a matter of fact, the function $g$
is well approximated by the Wigner-like distribution
\begin{equation}\label{fdist}
f(x)=\cN^{-1}y^\al\,\e^{-\frac{(y-\ga)^2}\be}\,,\qquad y=x+2\,,
\end{equation}
where the normalization constant $\cN$ is given by
\[
\cN=\frac12\,\al\be^{\frac\al2}\ga\Ga\Big(\frac\al2\Big)\ms
{}_1F_1\Big(\frac{1-\al}2\ms,\frac32\ms;-\frac{\ga^2}\be\Big)
+\frac12\,\be^{\frac{\al+1}2}\Ga\Big(\frac{\al+1}2\Big)\ms
{}_1F_1\Big(-\frac{\al}2\ms,\frac12\ms;-\frac{\ga^2}\be\Big),
\]
and ${}_1F_1$ is the confluent hypergeometric function of the
first kind. Thus the behavior of the chain $\cH$ is rather
different in this respect from that of the spin chains of HS type.

When $N$ is greater than $12$, the accuracy of the level density
obtained by the method of moments cannot be gauged directly by computing the
spectrum of the matrix $\cA$. An indirect way of estimating this accuracy
consists in comparing the first few moments
of the density $g$ with those of the spectrum of $\cA$,
computed by taking the traces of appropriate powers of $\cH$.
As an example, we shall next derive simple expressions for the mean and variance of the
spectrum of $\cH$ in terms of finite sums involving only the coordinates of the chain sites,
which can be easily evaluated for very large values of $N$. We shall see in this way that
the agreement between these values and those obtained from $g$ is indeed very good
and, roughly speaking, improves as $N$ grows.

\begin{figure}[h]
\begin{center}
\psfrag{I}[Bc][Bc][1][0]{\footnotesize $\log_{10}\De_i$}
\psfrag{N}{\footnotesize $N$}
\includegraphics[height=6cm]{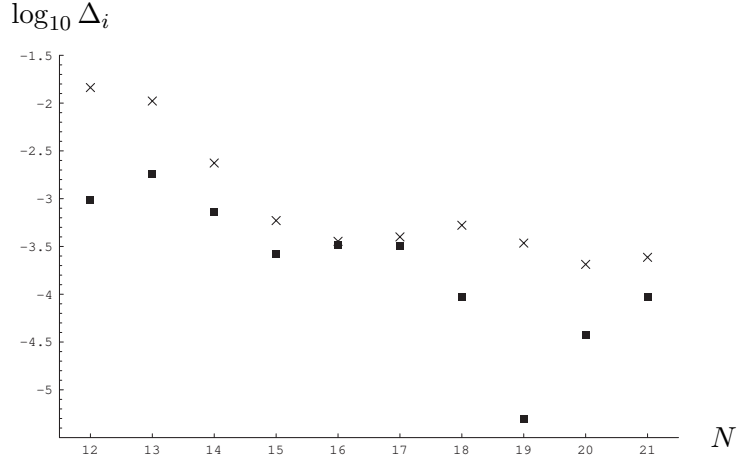}
\begin{quote}
  \caption{Logarithmic plot of the relative errors $\De_1$ (black boxes) and
    $\De_2$ (crosses) in Eq.~\eqref{relerrs} as a function of the number of
    sites $N$ for spin $1/2$. The approximate level density $g$ in
    Eq.~\eqref{Emsi2app} has been computed taking $k_0=20$ and averaging over
    $20$ random vectors.
  \label{fig:relerrs}}
\end{quote}
\end{center}
\end{figure}

In the first place, the mean energy $\BcE$ of $\cH$ can be easily computed noting that
$\tr S_{i,i+1}=n^{N-1}$, so that (cf.~Eq.~\eqref{hk})
\begin{equation}\label{cEmean}
\BcE=n^{-N}\tr\cH=n^{-N}\sum_i h_i(n^N-n^{N-1})=\Big(1-\frac1n\Big)\sum_i h_i\,.
\end{equation}
Similarly, since
\[
\cH^2=\sum_{i,j}h_ih_j(1-S_{i,i+1}-S_{j,j+1}+S_{i,i+1}S_{j,j+1})
\]
and
\[
\tr(S_{i,i+1}S_{j,j+1})=n^{N-2+2\de_{ij}}\,,
\]
a straightforward calculation yields
\[
\tr(\cH^2)=(n-1)n^{N-2}\Big((n-1)\big(\sum_i h_i\big)^2+(n+1)\sum_i h_i^2\Big),
\]
so that the variance of the energy is given by
\begin{equation}
  \si^2\equiv n^{-N}\tr(\cH^2)-{\BcE}^{\ms2}=\Big(1-\frac1{n^2}\Big)\sum_i h_i^2\,.
  \label{sigma2}
\end{equation}
In Fig.~\ref{fig:relerrs} we present a logarithmic plot of the relative errors
\begin{equation}
\De_1\equiv\frac{|\BcE-\BcE_{g}|}{\BcE}\,,\qquad
\De_2\equiv\frac{|\si^2-\si^2_{g}|}{\si^2}
\label{relerrs}
\end{equation}
between the exact values~\eqref{cEmean}-\eqref{sigma2} and their
approximations
\begin{equation}
  \BcE_{g}=\frac{\cE_{\max}}4\,\int_{-2}^2 (x+2) g(x)\d x\,, \quad
  \si^2_{g}=\frac{\cE_{\max}^2}{16}\,\int_{-2}^2 (x+2)^2 g(x)\d x
  -{\BcE}_{g}^{\ms2}
  \label{Emsi2app}
\end{equation}
for $N=12,\dots,21$ and spin $1/2$. From the latter plot it is
apparent that for $N\geq15$ both errors are less than $.1\%$,
which suggests that  for large $N$ the continuous function $g(x)$
computed by the method of moments is indeed an excellent
approximation to the level density of the matrix $\cA$.

{}From Eqs.~\eqref{cEmean} and~\eqref{sigma2} and the discussion
of the distribution of the chain sites in Section~\ref{S:Sites} it
is straightforward to deduce the asymptotic behavior of the mean
and the variance of the energy for large $N$. Indeed, using
Eq.~\eqref{dxdk} we easily obtain
\begin{equation}
h_k\equiv(\xi_{k+1}-\xi_k)^{-2}\simeq[x(k+1)-x(k)]^{-2}\simeq\Big(\frac{\d
x}{\d k}\Big)^{-2}\simeq\frac{N^2}{2\pi}\,\e^{-\xi_k^2}\,,
\end{equation}
so that
\begin{multline*}
\sum_k h_k\simeq\frac{N^2}{2\pi}\,\sum_k \e^{-\xi_k^2}
\simeq\frac{N^2}{2\pi}\,\sum_k \e^{-\xi_k^2}\cdot(\xi_{k+1}-\xi_k)\frac N{\sqrt{2\pi}}
\e^{-\frac{\xi_k^2}2}\\
\simeq\frac{N^3}{(2\pi)^{3/2}}\int_{-\infty}^\infty\e^{-\frac32\,x^2}\d x
=\frac{N^3}{2\pi\sqrt 3}\,.
\end{multline*}
Hence for large $N$ the mean energy $\BcE$ behaves as
\begin{equation}\label{largeBcE}
\BcE\simeq\frac{N^3}{2\pi\sqrt 3}\,\Big(1-\frac1n\Big)\,.
\end{equation}
Similarly,
\[
\sum_k h_k^2
\simeq\frac{N^4}{4\pi^2}\,\sum_k \e^{-2\xi_k^2}\cdot(\xi_{k+1}-\xi_k)\frac N{\sqrt{2\pi}}
\e^{-\frac{\xi_k^2}2}
\simeq\frac{N^5}{(2\pi)^{5/2}}\int_{-\infty}^\infty\e^{-\frac52\,x^2}\d x
=\frac{N^5}{4\pi^2\sqrt5},
\]
and therefore the large $N$ limit of the variance of the energy is given by
\begin{equation}\label{largesi}
\si^2\simeq\frac{N^5}{4\pi^2\sqrt5}\,\Big(1-\frac1{n^2}\Big)\,.
\end{equation}
Thus, the leading behavior of the energy mean and variance of the chain~\eqref{cH}
is analogous to that of the trigonometric HS spin
chains studied in Refs.~\cite{FG05,EFGR05}. In particular, both quantities
grow with $N$ much faster than their counterparts for the
Heisenberg chain (both of which diverge as $N$).

\section{Conclusions}\label{S:conc}

In this paper we have studied a novel spin chain with position-dependent
nearest neighbors interactions, which is intermediate between the Heisenberg chain
(position-independent, nearest neighbors interactions) and the spin chains
of Haldane--Shastry type (position-dependent, long-range interactions);
see Table~\ref{table:properties} for a brief comparison.

We have developed a new method related to Polychronakos's freezing trick which
has made it possible to compute in closed form a certain number of energy
levels with their corresponding eigenstates, for any number of particles and
arbitrary spin. While the eigenvalues of the original Haldane--Shastry and
Polychronakos--Frahm chains (the closest HS analogs of the chain under
consideration) are all integers, the only integer eigenvalues of our chain are
precisely those that have been computed exactly by our method. This fact
strongly suggests that the chain~\eqref{cH} should be regarded as the first
instance of a quasi-exactly solvable spin chain, thus extending the usual
notion of quasi-exact solvability to finite-dimensional Hamiltonians.
\begin{table}[h]
\caption{A comparison of some properties of the spin chain $\cH$
versus the Heisenberg chain $\cH_{\mathrm{He}}$ and the
Polychronakos--Frahm chain~$\cH_{\mathrm{PF}}$ (the properties followed
by an asterisk are based on unpublished work by the authors).\vskip2mm}\label{table:properties}
\begin{center}
\begin{tabular}{lccc}\hline
  \vrule height 11pt depth 7pt width0pt & $\cH$ &
  $\cH_{\mathrm{He}}$ & $\cH_{\mathrm{PF}}$\\\hline
Chain sites distribution & Gaussian & Equispaced & Circular law \eqref{rhoPF}\\ \hline
Solvability & Quasi-exact & Exact & Exact\\ \hline
Integer energies & First few & No & All\\ \hline
Level density & Wigner-like law \eqref{fdist} & Wigner-like${}^{\ms*}$ & Gaussian${}^{\ms*}$\\ \hline
Mean energy growth & $N^3$ & $N$ & ${N^2}^{\,*}$\\ \hline
Energy variance growth & $N^5$ & $N$ & ${N^3}^{\,*}$\\ \hline
\end{tabular}
\end{center}
\end{table}

The method developed in this paper is quite general, and only relies on the explicit
knowledge of a number of eigenfunctions of the corresponding spin dynamical model.
In this respect, it goes one step beyond the usual freezing trick, which requires the
computation in closed form of the whole spectrum of the related dynamical model.
In particular, our method is well-suited to spin chains whose associated
dynamical model is only quasi-exactly solvable, like for instance
the models with elliptic interactions constructed in Refs.~\cite{FGGRZ01,FGGRZ01b}.

\begin{ack}
This work was partially supported by the DGI under grant
no.~FIS2005-00752, and by the Complutense University and the
DGUI under grant no.~GR69/06-910556. A.E. acknowledges the financial support of the
Spanish Ministry of Education and Science through an FPU
scholarship. The authors would also like to thank
V.~Mart{\'\i}n-Mayor for useful discussions on the moments method.
\end{ack}

\begin{appendix}
\section{Exact eigenfunctions of the dynamical spin model~\eqref{H}}\label{appA}

In this appendix we shall list the eigenfunctions of the dynamical spin model~\eqref{H}
used to construct the eigenstates of the chain~\eqref{cH} presented in Section~\ref{S:QES}.
Let
\[
\al=N\Big(a+\frac12\Big)-\frac32\,,\qquad
\be\equiv\be(m)=1-m-N\Big(a+\frac12\Big)\,,\qquad
t=\frac{2r^2}{N\Bx^2}-1\,,
\]
where $\Bx=\frac 1N\sum_ix_i$ is the center of mass coordinate.
In Ref.~\cite{EFGR07} it was shown that the model~\eqref{H} possesses
the following families of spin eigenfunctions with energy $E_{lm}=E_0+2a(2l+m)$,
where $l\geq 0$ and $m$ is as indicated in each case:
\begin{align}
\Psi^{(0)}_{lm}&=\mu_0\ms\Bx^mL^{-\be}_l(a r^2)
P^{(\al,\be)}_{\lf\frac m2\rf}(t)\,\Phi^{(0)}\,,\qquad m\geq0\,,\label{Psi0}\\
\Psi^{(1)}_{lm}&=\mu_0\ms\Bx^{m-1}L^{-\be}_l(a
r^2)P^{(\al+1,\be)}_{\lf\frac{m-1}2\rf}(t)\big(\Phi^{(1)}
-\Bx\,\Phi^{(0)}\big)\,,\qquad m\geq1\,,\displaybreak[0]\label{Psi1}\\
\Psi^{(2)}_{lm}&=\mu_0\ms\Bx^{m-2}L^{-\be}_l(a
r^2)\bigg[P^{(\al+2,\be)}_{\lf\frac m2\rf-1}(t)\big(\Phi^{(2)}-2\Bx\,\Phi^{(1)}+\Bx^2\,\Phi^{(0)}\big)\notag\\
&\hspace{8em}-\frac{2(\al+1)}{2\lf\tfrac{m-1}2\rf+1}\,\Bx^2P^{(\al+1,\be)}_{\lf\frac
m2\rf-1}(t)\,\Phi^{(0)}\bigg]\,,\quad m\geq2\,,\displaybreak[0]\label{Psi2}\\
\widetilde\Psi^{(2)}_{lm}&=\mu_0\ms\Bx^{m-2}L^{-\be}_l(a
r^2)\bigg[P^{(\al+2,\be)}_{\lf\frac m2\rf-1}(t)\big(\widetilde\Phi^{(2)}-2\Bx\,\Phi^{(1)}
+\Bx^2\Phi^{(0)}\big)\notag\\
&\hspace{4em}+\frac{2(\al+1)}{\big(2\lf\tfrac{m-1}2\rf+1\big)(N-1)}\,\Bx^2P^{(\al+1,\be)}_{\lf\frac
m2\rf-1}(t)\Phi^{(0)}\bigg]\,,\quad m\geq2\,,\displaybreak[0]\label{tPsi2}\\
\hPsi^{(3)}_{lm}&=\mu_0\ms\Bx^{m-3}L^{-\be}_l(a r^2)\bigg[
P^{(\al+3,\be)}_{\lf\frac{m-3}2\rf}(t)\big(\hPhi^{(3)}-2\Bx\,\Phi^{(2)}
+2\Bx^2\,\Phi^{(1)}-\frac23\,\Bx^3\Phi^{(0)}\big)\notag\\
&\hspace{6em}+\frac{2\ms\Bx^2}{2\lf\frac m2\rf-1}\,P^{(\al+2,\be)}_{\lf\frac{m-3}2\rf}(t)
\big(2(\al+2)\,\Phi^{(1)}-\Bx\Phi^{(0)}\big)\notag\\
&\hspace{6em}-\frac{2(2\al+3)}{m(m-2)}\,\vep(m)\,\Bx^3
P^{(\al+1,\be)}_{\frac{m-3}2}(t)\,\Phi^{(0)}\bigg]
\,,\quad m\geq3\,.\displaybreak[0]\label{hPsi3}
\end{align}
In the previous formulas $\lf\cdot\rf$ denotes the integer part and
$\vep(m)=\frac12(1-(-1)^m)$ is the parity of $m$.
The spin functions $\Phi^{(k)}$, $\tPhi^{(2)}$, $\hPhi^{(3)}$ (cf.~Eqs.~\eqref{Phis}
and~\eqref{hPhi}) are built from a
state $\ket s$ symmetric under $S_{12}$ and belonging to $\Si'$
for $\widetilde\Psi^{(2)}_{lm}$, and antisymmetric under $S_{12}$
for $\hPsi^{(3)}_{lm}$. The generalized Laguerre polynomials $L^{-\be}_l$
and the Jacobi polynomials $P^{(\ga,\be)}_k$ appearing in Eqs.~\eqref{Psi0}--\eqref{hPsi3}
are defined as
\begin{gather}
L^{-\be}_l(z)=\sum_{j=0}^l(-1)^j\binom{l-\be}{l-j}\frac{z^j}{j!}\,,\\
P^{(\ga,\be)}_k(z)=\frac1{k!}\,\sum_{j=0}^k\frac1{2^jj!}\,
(-k)_j(\ga+\be+k+1)_j(\ga+j+1)_{k-j}\,(1-z)^j\,,\label{P}
\end{gather}
where
\[
(x)_j=x(x+1)\cdots(x+j-1)
\]
is the Pochhammer symbol. The eigenfunctions of type~\eqref{hPsi3} are independent
of the remaining ones~\eqref{Psi0}--\eqref{tPsi2} only for spin greater than $1/2$ ($n>2$),
cf.~\cite{EFGR07}. The model~\eqref{H} possesses two additional families of eigenfunctions
(also derived in the previous reference),
which have not been listed above as they do not yield any further eigenstates
of the chain~\eqref{cH}.

Let us now discuss the behavior of the terms $L^{-\be}_l(ar^2)$
and $\Bx^{2k}P^{(\al+i,\be)}_k(t)$ in Eqs.~\eqref{Psi0}--\eqref{hPsi3}
as $a\to\infty$. Consider first the polynomial $L^{-\be}_l(ar^2)$. Since
\[
\binom{l-\be}{l-j}=\frac{(Na)^{l-j}}{(l-j)!}\,\big(1+O(a^{-1})\big)\,,
\]
it follows that
\begin{equation}\label{La}
a^{-l}L^{-\be}_l(ar^2)=\frac{(N-r^2)^l}{l!}+O(a^{-1})\,,
\end{equation}
where the term $O(a^{-1})$ in the previous equation is actually a polynomial in $a^{-1}$.
(Throughout this paper, the symbol $O(a^{-k})$ denotes any function $f(a)$ such that
$a^k f(a)$ has a finite (possibly zero) limit as $a\to\infty$.)
For the other type of term, note that
\begin{multline}\label{BxP}
\Bx^{2k}P^{(\al+i,\be)}_k(t)
=\frac1{k!}\sum_{j=0}^k\bigg[\frac{(-k)_j}{N^jj!}(\al+\be+i+k+1)_j\\
\times(\al+i+j+1)_{k-j}\,\Bx^{2(k-j)}{(N\Bx^2-r^2)}^j\bigg]
\end{multline}
is clearly a polynomial in $\bx$. Taking into account that
\[
(\al+i+j+1)_{k-j}=(Na)^{k-j}\,\big(1+O(a^{-1})\big)\,,
\]
we obtain
\begin{equation}\label{Pa}
a^{-k}\Bx^{2k}P^{(\al+i,\be)}_k(t)=\frac{N^k}{k!}\,\Bx^{2k}+O(a^{-1})\,,
\end{equation}
where $O(a^{-1})$ is polynomial in $a^{-1}$ and $\bx$.

\section{The method of moments}\label{appB}

The method of moments~\cite{BRP92,AFGLM01} is a powerful tool for computing
the density of eigenvalues of a large Hermitian matrix whose spectrum
is not known explicitly. The method is based on the relation between
a probability distribution $g(x)$, i.e., a nonnegative function
whose integral over its support $[x_1,x_2]$ is one, and its {\em resolvent}
\[
R(z)=\int_{x_1}^{x_2}\frac{g(x)}{z-x}\,\d x\,,\qquad \Imag z\neq0\,.
\]
Using the well-known identity
\[
\lim_{\ep\to0+}\frac{\ep}{\ep^2+x^2}=\pi\de(x)\,,
\]
it is straightforward to show that
\begin{equation}\label{g}
g(x)=\mp\frac1\pi\,\lim_{\ep\to0+}R(x\pm\iu\ms\ep)\,,\qquad x\in(x_1,x_2)\,.
\end{equation}
The previous formula makes it possible to compute the probability
density $g(x)$ if the resolvent $R(z)$ is known. One of the key
ingredients of the method is the fact that if $g$ is positive in a
set of nonzero measure, the resolvent can be expanded as a
continued fraction
\begin{equation}\label{contf}
R(z)=\cfrac{1}{z-b_0-\cfrac{c_1}{z-b_1-\dotsb}}\;,
\end{equation}
where $b_k$, $c_k$ are the coefficients in the three-term recursion relation
\begin{equation}\label{3term}
P_{k+1}(x)=(x-b_k)P_k(x)-c_kP_{k-1}(x)\,,\qquad
(P_{-1}=0\,,\; P_0=1\,,\; c_0=1)
\end{equation}
satisfied by the orthogonal polynomial system ${\{P_k(x)\}}_{k=0}^\infty$ associated
with the density $g(x)$. The polynomials $P_k$ are the monic polynomials
determined by the orthogonality condition
\[
\langle P_kP_l\rangle\equiv\int_{x_1}^{x_2}P_k(x)P_l(x)g(x)\d x=0\,,\qquad k\neq l\,.
\]
It can be shown~\cite{Ch78} that these polynomials satisfy a three-term recursion relation of the
form~\eqref{3term} with coefficients $b_k$ and $c_k$ given by
\begin{equation}\label{bkck}
b_k=\frac{\langle xP_k^2\rangle}{\langle P_k^2\rangle}\,,\qquad
c_k=\frac{\langle P_k^2\rangle}{\langle P_{k-1}^2\rangle}\,.
\end{equation}

In this paper we are interested in computing the
density of eigenvalues (normalized to unity) of a Hermitian $d\times d$ matrix $\cA$, i.e.,
\[
g(x)=\frac1d\,\sum_{i=1}^d\de(x-\la_i)\,,
\]
where $\la_1\leq\dots\leq\la_d$ are the eigenvalues of $\cA$. The corresponding resolvent
\[
R(z)=\frac1d\,\sum_{i=1}^d(z-\la_i)^{-1}=\frac1d\,\tr\ms(z-\cA)^{-1}
\]
is thus a rational function of $z$. In this case the resolvent can
also be expanded as a continued fraction of the form~\eqref{contf}
terminating at level $d$, i.e., $c_d=0$ and Eqs.~\eqref{3term}
and~\eqref{bkck} hold for $k\leq d-1$.

If the density $g$ is not known, the coefficients $b_k$ and $c_k$ cannot be
computed from Eqs.~\eqref{bkck}, and hence the expansion~\eqref{contf} cannot
be used directly to evaluate the resolvent. On the other hand, the
coefficients $b_k$ and $c_k$ can be determined once the moments $\langle x^k\rangle$ of
the distribution $g$ are known. Another central idea of the method consists in
replacing $\langle x^k\rangle$ by the average of the expectation values $(\bv,\cA^k\bv)$
over a suitable set of normalized random vectors~$\bv\in\RR^d$. More
precisely, let $u_i$, $i=1,\dots,d$, be $d$ independent random variables
uniformly distributed in the interval $[-1,1]$, and let $\bv$ be the vector
with components $v_i=u_i/\|\bu\|$. Clearly, the components of $\bv$ satisfy
\begin{equation}
\overline{v_iv_j\vphantom{t}}=\frac1d\,\de_{ij}\,,
\end{equation}
where the bar stand for the average over the random numbers $u_i$.
We thus have
\begin{align}
\overline{(\bv,\cA^k\bv)\vrule height10pt width0pt}&=\sum_{i,j=1}^d(\cA^k)_{ij}\,\overline{v_iv_j\vphantom{t}}
=\frac1d\,\sum_{i=1}^d(\cA^k)_{ii}=\frac1d\,\tr(\cA^k)
=\frac1d\,\sum_{i=1}^d\la_i^k\notag\\
&=\frac1d\,\sum_{i=1}^d\int x^k\de(x-\la_i)\d x=\langle x^k\rangle\,.
\end{align}
The previous equality implies that
\[
\langle P(x)\rangle=\overline{(\bv,P(\cA)\bv)\vrule height10pt width0pt}\,,
\]
where $P$ is an arbitrary polynomial. {}From Eq.~\eqref{bkck} we immediately
obtain the following formula for the coefficients $b_k$ and $c_k$:
\begin{equation}\label{bkckrand}
b_k=\frac{\,\overline{(\bw_k,\cA\bw_k\vrule height10pt width0pt)}\,}
{\vrule height13pt width0pt\overline{\|\bw_k\|^2\vrule height10pt width0pt}}\,,\qquad
c_k=\frac{\vrule height13pt width0pt\overline{\|\bw_k\|^2\vrule height10pt width0pt}}
{\,\vrule height13pt width0pt\overline{\|\bw_{k-1}\|^2\vrule height10pt width0pt}\,}\,,
\end{equation}
where $\bw_k\equiv P_k(\cA)\bv$.
By Eq.~\eqref{3term}, the vectors $\bw_k$ satisfy the recursion relation
\[
\bw_{k+1}=(\cA-b_k)\bw_k-c_k\bw_{k-1}\qquad
(\bw_{-1}=0\,,\; \bw_0=\bv\,,\; c_0=1).
\]
The latter formula, together with Eq.~\eqref{bkckrand}, can be used
to recursively compute the coefficients $b_k$ and $c_k$ for all $k=1,\dots,d-1$.

The number $d=n^N$ is typically very large, so that it is not
feasible to compute all the coefficients $b_k$ and $c_k$ by the
procedure just outlined. In practice, one only computes the first
few of these coefficients, say up to $k=k_0$, and tries to
estimate the part $T_{k_0}(z)$ of the continued
fraction~\eqref{contf} involving the remaining coefficients. In
many cases of interest~\cite{TDT82}, the coefficients $b_k$ and $c_k$
approximately stabilize for a wide range of values of $k$, namely
$b_k\simeq b$ and $c_k\simeq c$ for $k_0<k<k_1<d$, where $k_1\gg
k_0$. In this case the remainder $T_{k_0}(z)$ can be approximated
as follows:
\[
T_{k_0}(z)
\simeq\cfrac{c}{z-b-\cfrac{c}{z-b-\dotsb}}
\equiv T(z)\,.
\]
But the truncation factor $T(z)$ satisfies
\[
T(z)=\frac c{z-b-T(z)}\,,
\]
so that
\begin{equation}\label{T}
T(z)=\frac12\big(z-b\pm\sqrt{(z-b)^2-4c}\big)\,.
\end{equation}
We thus obtain the following approximate formula for the
resolvent:
\begin{equation}\label{contfT}
R(z)\simeq\cfrac{1}{z-b_0-\cfrac{c_1}{z-b_1-\;\lower 10pt\hbox{$\ddots$}
\lower 20pt\hbox{$\:{}-\cfrac{c_{k_0}}{z-b_{k_0}-T(z)}$}}}\;.
\end{equation}
The previous equation can be written as
\begin{equation}\label{R}
R(z)\simeq\frac{A(z)\pm\iu B(z)\sqrt{\om(z)}}{C(z)\pm\iu D(z)\sqrt{\om(z)}}\,,
\end{equation}
with $A,B,C,D$ polynomials in $z$ with real coefficients and
\begin{equation}\label{om}
\om(z)=(z-x_1)(x_2-z)\,,\qquad x_{1,2}=b\pm2\sqrt c\,;
\end{equation}
note that $x_{1,2}\in\RR$, on account of Eq.~\eqref{bkckrand}. The
RHS of~\eqref{R} has a branch cut on the interval $[x_1,x_2]$,
which is the support of its corresponding density function.
{}From Eqs.~\eqref{g} and~\eqref{R} we finally obtain the following
approximate formula for the normalized density of eigenvalues of
the Hermitian matrix~$\cA$:
\begin{equation}\label{gfinal}
g(x)\simeq\frac{\vert A(x)D(x)-B(x)C(x)\vert}{C(x)^2+D(x)^2\om(x)}\,\frac{\sqrt{\om(x)}}\pi\,,
\qquad x\in[x_1,x_2]\,.
\end{equation}
Since the RHS of the previous equation cannot vanish identically in any
subinterval of $[x_1,x_2]$ with nonzero length, it follows that the
formula just obtained is only appropriate when the spectrum
of $\cA$ has no gaps.
\end{appendix}


\end{document}